\documentclass[aps,prb,superscriptaddress,,twocolumn]{revtex4}
\usepackage{amsmath}
\usepackage{amssymb}
\usepackage{amsthm}
\usepackage{graphicx}
\usepackage{graphics}
\usepackage{bm}
\usepackage{hyperref}
\usepackage{rotating}
\usepackage{color}
\newcommand{\bra}[1]{\langle #1|}
\newcommand{\ket}[1]{|#1\rangle}

\def\beas{\begin{eqnarray*}}
\def\eeas{\end{eqnarray*}}
\def\bea{\begin{eqnarray}}
\def\eea{\end{eqnarray}}
\def\be{\begin{equation}}
\def\ee{\end{equation}}
\newcommand{\sg}{\sigma}
\newcommand{\ua}{\uparrow}
\newcommand{\da}{\downarrow}

\newcommand{\D}{\Delta}

\begin{document}

\title{Experimental signatures of non-Abelian statistics
        in clustered quantum Hall states}%

\author{Roni Ilan}
\affiliation{Department of Condensed Matter Physics, Weizmann
Institute of Science, Rehovot 76100, Israel}

\author{Eytan Grosfeld}
\affiliation{Department of Condensed Matter Physics, Weizmann
Institute of Science, Rehovot 76100, Israel}

\author{Kareljan Schoutens}
\affiliation{Institute for Theoretical Physics, University of Amsterdam, Valckenierstraat 65, 1018 XE Amsterdam, the Netherlands}

\author{Ady Stern}
\affiliation{Department of Condensed Matter Physics, Weizmann
Institute of Science, Rehovot 76100, Israel}

\date{March 10, 2008}

\begin{abstract}
We discuss transport experiments for various non-Abelian quantum
Hall states, including the Read-Rezayi series and a paired spin
singlet state. We analyze the signatures of the unique characters
of these states on Coulomb blockaded transport through large
quantum dots. We show that the non-Abelian nature of the states
manifests itself through modulations in the spacings between
Coulomb blockade peaks as a function of the area of the dot. Even
though the current flows only along the edge, these modulations
vary with the number of quasiholes that are localized in the bulk
of the dot. We discuss the effect of relaxation of edge states on
the predicted Coulomb blockade patterns, and show that it may
suppress the dependence on the number of bulk quasiholes. We
predict the form of the lowest order interference term in a
Fabry-Perot interferometer for the spin singlet state. The result
indicates that this interference term is suppressed for certain
values of the quantum numbers of the collective state of the bulk
quasiholes, in agreement with previous findings for other
clustered states belonging to the Read-Rezayi series.
\end{abstract}


\maketitle


\section{introduction}\label{sec:introduction}

Probing  exotic quantum statistics of particles is a long standing
experimental challenge. In fractional quantum Hall effect systems,
elementary excitations are expected to follow fractional statistics.
There, it is difficult to experimentally distinguish the effects of
Abelian fractional statistics from the effects associated with the
fractional charge attributed to the same excitations. However, for
states in which the statistics is expected to be non-Abelian, it is
predicted that once the correct signature of this property is
identified and measured, there will be no ambiguity in associating
it with quantum statistics. For this reason alone, it is worth
making the effort to explore the implication of non-Abelian
statistics on measurable quantities.

Non-Abelian statistics is a key ingredient in the design of quantum
gates for topologically protected qubits
\cite{Kitaev,sarma,TQCrev,SternAnyonReview}. Experimental proof of
the existence of non-Abelian anyons would be a major step towards an
implementation of topological quantum computation.

Most suggested experiments aimed at probing non-Abelian statistics
in quantum Hall systems consider a system designed to form a two
path interferometer. The path difference between two quasiparticle
(or quasihole) trajectories along the edge of the system forms a
closed loop around a part of the system. The electronic density is
determined by the positive background charge density and does not
vary with magnetic field. Hence, a small deviation in the external
applied magnetic field, introduces quasiparticles  or quasiholes
(depending on whether the magnetic field is increased or decreased),
localized by impurities, into the loop
\cite{Fradkin,SternHalperin,Bonderson1,Bonderson2,Stone}. As the
number of these quasiparticles/quasiholes, $n$, is varied, the two
terminal conductance is found to show a strong $n$-dependence via an
interference term. Other experiments\cite{SternHalperin,ilan}
considered tunneling of electrons into a large quantum dot whose
interior is in a gapped quantum Hall state, and predicted how the
clustering property of the Read-Rezayi (RR) states should influence
Coulomb blockade peaks.

In this paper we have two goals. First, we extend the discussion
of transport through dots in quantum Hall states belonging to the
Read-Rezayi series and characterized by Coulomb blockade; we
particularly address the issue of the different time scales
involved in the experiment. The fastest time scale is the time in
which a current carrying electron traverses the dot, $e/I$. For
these electrons the bulk of the dot is inaccessible, and they
occupy the lowest energy state on the edge. A much slower time
scale is that in which the area of the dot is varied. If this
scale is slow enough, the electrons that are added to the dot when
its area is varied may make use of bulk states. We assume, for
concreteness, that the magnetic field is tuned such that there are
quasiholes localized in the bulk, rather than quasiparticles. The
internal bulk states are then carried by these localized
quasiholes. For that to happen, however, weak coupling should
exist between the bulk quasiholes and the edges.

In previous discussions of Coulomb blockade measurements in such an
experiment\cite{SternHalperin,ilan}, the hidden assumption was that
as the variation of the area adds electrons to the dot, these added
electrons make use only of states at the edge.  In the language of
topological field theory (TFT), the fusion channel of all the
quasihole operators in the bulk of the dot is fixed at the beginning
of the experiment, and does not change as new electrons are added to
the dot. In the present paper we consider the possibility that as
the area is slowly varied, weak coupling of the bulk and the edge
may allow for the fusion channel to change when electrons are added,
in such a way that energy is minimized.

Second, we analyze the Fabry-Perot setting in another non-Abelian
quantum Hall state, a $\nu=4/7$ spin-singlet state, whose
experimental signatures have not been considered so far. This
state was suggested by Ardonne and Schoutens\cite{first-ardonne}.
It is the first in a series of spin-singlet states with order-$k$
clustering, at filling fraction $\nu=\frac{2k}{2k+3}$. The $k>1$
states support spinful quasihole excitations with non-Abelian
statistics. For the paired ($k=2$) state at $\nu=4/7$ the
quasiholes are Fibonacci anyons \cite{ardonne-2007-322}.
Experimentally, a spin transition for quantum Hall states around
$\nu=4/7$ has been observed in samples with reduced Zeeman
splitting \cite{choetal}, but this feature may also indicate an
unpolarized Abelian state of the Jain series. Transport properties
we analyze here may distinguish between the two candidate states.
The fact that possible fractional quantum Hall states in graphene
are expected to be spin-singlet with respect to the pseudo-spin
(valley) index \cite{Graphene}, provides additional motivation for
studying non-Abelian statistics for spin-singlet states. In this
paper we study Coulomb blockade transport through a dot in the
$\nu=4/7$ non-Abelian spin-singlet state, with and without
relaxation. We also analyze the lowest order interference in a
Fabry-Perot interferometer and derive a characteristic suppression
factor.

The paper is organized as follows. In Sec. \ref{sec:experimental},
we discuss the experimental setup in which these experiments are to
be carried out. In Sec. \ref{sec:CFT intro} we discuss in short the
general structure of the conformal field theories (CFT) describing
these states. In Sec. \ref{sec:RR and parafermions} we discuss in
detail the formation of Coulomb blockade peaks in the conductance
through large quantum dots in the RR states. In Sec. \ref{sec:AS
states} we turn to implement the same ideas to the spin singlet
state in order to predict the result for the Coulomb blockade peaks.
We also calculate the interference term of the current in a two path
interferometer for this state.

\section{General considerations}\label{sec:experimental}

\begin{figure}[t]
\includegraphics[width=0.17\textwidth,angle=-90]{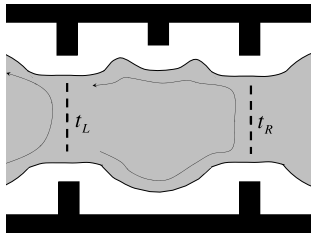}
\caption{Setup for an interference experiment. The gray region is
a gapped quantum Hall fluid. Two possible tunneling paths for an
incoming edge quasihole are marked. There are $n$ quasiholes
localized in the area between the two quantum point contacts.}

\label{fig:inter-weak}
\end{figure}\

\begin{figure}
\includegraphics[width=0.25\textwidth]{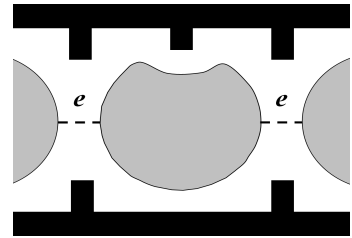}
\caption{Fabry Perot interferometer in the limit of strong
quasihole backscattering. Transport through the dot is done via
electron ($e$) tunneling. The area $S$ of the dot may be varied
using a side modulation gate.} \label{fig:inter-strong}
\end{figure}

The Fabry-Perot interferometer, sketched in
Fig.~\ref{fig:inter-weak}, is a Hall bar with two quantum point
contacts (QPCs) introducing quasiparticle/quasihole tunneling from
one edge to the other. We deal with two opposite limits of this
interferometer: weak inter-edge back-scattering, in which we look at
interference to lowest order, and strong inter-edge backscattering,
where the interferometer becomes a quantum dot, whose Coulomb
blockade peaks we study.

As mentioned above, we shall always assume that there are
quasiholes, rather than quasiparticles,  that are localized in the
bulk. For the purpose of lowest order interference calculations,
when one is also required to consider excitations on the edge, we
shall consider for convenience that the current along the edge is
also carried by quasiholes. Changing this current to be of
quasiparticles should not modify our results.

Lowest order interference is observed when a single quasihole does
not tunnel between opposite edges more than once. The tunneling
process introduces a finite value for the longitudinal conductance.
The measured backscattered current, injected from the left along the
lower edge and collected on the left at the upper edge, interferes
only through two trajectories: quasiholes entering from the left
along the lower edge are either being backscattered at the left QPC,
or transmitted at the left QPC and reflected from the right one. The
path difference between the two trajectories forms a closed loop
around the island confined between the contacts, which may contain
localized quasiholes in the bulk of the sample. For the sake of
discussing lowest order interference, we assume that there is no
hopping of quasiholes between the bulk and the edge.

The two  trajectories sketched in Fig.~\ref{fig:inter-weak} are
associated with two partial waves, $\ket{\Psi_L}$ and
$\ket{\Psi_R}$. While these partial waves differ by global phases
originating from the total magnetic field through the island (due
to the Aharonov-Bohm effect), their overlap also encodes
information on the mutual statistics of the quasiholes in the
system. As mentioned in sec.~\ref{sec:introduction}, this is
attributed to the fact that the path difference between the two
trajectories of the edge qusiholes forms a closed loop encircling
localized quasiholes. Altogether, the back-scattered current will
be of the form
\begin{equation}\label{eq:sigma xx}
I_{bs}\propto |t_L|^2+|t_R|^2+ 2 \,
\text{Re}\left\{t_L^*t_R\bra{\psi_L}\psi_R\rangle\right\}.
\end{equation}
It is the overlap $\bra{\psi_L}\psi_R\rangle$ that we calculate
below.

In the limit of strong back-scattering
(Fig.~\ref{fig:inter-strong}), when the two point contacts on either
side of the island are almost closed, the number of electrons in the
dot confined between them is quantized to an integer. Quasihole or
quasiparticle tunneling into and out of this region is forbidden,
and the only way to transport charge through it is by tunneling of
electrons. Low-voltage low-temperature conductance through the dot
is suppressed due to charging energy, except at "Coulomb blockade
peaks", when the ground state energy of the dot with $N_e$ electrons
is degenerate with its ground state energy with $N_e+1$ electrons.
Coulomb blockade peaks may be probed by measuring the conductance
through the dot as a function of a magnetic field $B$ and the dot's
area $S$, since a variation of $S$ with fixed $N_e$ violates charge
neutrality with the positive background \cite{ilan}. Peaks in the
conductance appear for those values of the area $S$ and magnetic
field $B$, for which the following equation
\begin{equation}
E(N_e,S,B)=E(N_e+1,S,B) \label{coulomb-peaks}
\end{equation}
is satisfied for some integer $N_e$ representing the total number of
electrons inside the dot.

The quantum Hall liquid is largely gapped, with two exceptions:
the edge and the internal degrees of freedom of the bulk
quasiholes. As we will review below, the state of the internal
degrees of freedom  of the bulk quasiholes determines the spectrum
of the edge, that determines, in turn, the position of the Coulomb
peaks.

In Ref.~(\onlinecite{ilan}), Coulomb blockade peaks were mapped for
large dots  in a quantum Hall state of the RR series
$\nu=\frac{k}{k+2}$, under the assumption that the state of the bulk
quasiholes is frozen. For a clean large ($N_e \gg 1$) dot in a
metallic state at zero magnetic field, and dots in the integer and
Abelian fractional quantum Hall states the area spacings between
consecutive Coulomb blockade peaks is $\Delta S=e/n_0$, the area
occupied by one electron. In contrast, for the RR series the Coulomb
blockade peaks location as a function of the area at a fixed
magnetic field (i.e., a fixed number of localized bulk quasiholes
$n$) was found to depend on $B$. While the average spacing between
peaks remains $e/n_0$, the presence of non-Abelian quasiholes in the
bulk causes the peaks to bunch into groups, where the number of
peaks in each group depends on $k$ and on $n$.

The edge energies of the non-Abelian quantum Hall states we discuss
in this paper are stored in a bosonic charged edge mode (a chiral
Luttinger liquid) and one or several neutral edge modes. With
$N_e=0$ defined to be the number of electrons for a dot with area
$S_0$, the energy associated with the charge mode for $N_e$
electrons on the edge is \cite{ilan}
\begin{equation}\label{eq:charging}
E_c(N_e)=\frac{\pi v_c}{\nu L}\left(N_e-\nu\frac{B_{0}(
S-S_0)}{\phi_0}\right)^2.
\end{equation}
where $\phi_0$ is the magnetic flux quantum. The magnetic field
$B=B_{0}$ is that in which there are no quasiholes in the bulk,
$\nu$ is the filling fraction of the partially filled topmost
Landau level, $v_c$ is the velocity of the charged edge mode, and
$L$ is the perimeter of the quantum dot (we take $\hbar=1$). For
Abelian Laughlin states, where there is only a single bosonic
mode, Eq.~(\ref{coulomb-peaks}) reduces to $E_c(N_e)=E_c(N_e+1)$,
and the area separation $\Delta S$ between its solutions for
consecutive values of $N_e$ is $\Delta S=e/n_0$, where $n_0$ is
the charge density inside the dot. The value of $\Delta S$ in this
case is independent of the magnetic field. For non-Abelian states,
this is not the case. Below, we focus on the contribution of the
neutral mode, and add the charge contribution
Eq.~(\ref{eq:charging}) at the final stage.

The assumptions underlying the calculation performed in
Ref.~(\onlinecite{ilan}) were that the magnetic field, and therefore
the number of quasiholes, is fixed at the beginning of the
experiment, and that the dot is relaxed into its ground state. It
was also assumed that the initial number of electrons inside the dot
was divisible by $k$, such that all electrons are clustered in the
bulk of the dot and the occupation of the electron states of the
edge is zero. Then, with a fixed number of quasiholes in the bulk,
the area is varied fast enough such that there is no time for the
electrons to relax onto states with lower energies that may be
available.

Here, we allow the initial number of electrons to have any integer
value. For concreteness we assume that the dot has no bulk
quasiholes in the beginning of the experiment. The number of
electrons then dictates the  number of edge modes that are
initially occupied. Shifting the magnetic field creates a fixed
number of quasiholes in the bulk, whose fusion channel is
determined by energetic considerations we explain in details later
on. The process of introducing quasiholes into the bulk is done
slowly enough for the edge and the bulk to equilibrate. Finally,
once the number of quasiholes and their fusion channel is
determined, we move on to consider what happens when the area of
the dot is varied.

In our analysis we consider the implications of varying the area
slowly enough such that the state of the bulk quasiholes may
changes adiabatically. As the area of the dot grows, more
electrons are added to the dot. In the absence of edge-bulk
coupling, the state of the bulk quasiholes does not change, and
the added electrons affect only the state of the edge. With
bulk-edge coupling, tunneling of neutral modes between the edge
and the bulk quasiholes may change the state of the bulk
quasiholes.

We may envision both elastic and inelastic bulk-edge couplings. In
the former, a tunnel coupling allows for the hopping of a neutral
particle (whose properties are to be described
below\cite{BulkEdgeCoupling,OverboschWen}) between the bulk and
the edge. In the latter, this hopping is accompanied by an energy
transfer to an outside thermal bath, e.g., by an emission of a
phonon. The latter is thus also an irreversible relaxation
mechanism that allows the system to cool. Quantitative estimate of
these two types of couplings are not possible at our present level
of understanding of the microscopy of the samples and the involved
quantum Hall states. Both are, however, bound to exist to some
level. In this work we study in details the process involving
energy relaxation. We also discuss qualitatively the effect of the
first process for the particular case of $\nu=5/2$.

The following argument may illuminate the way the state of the
bulk quasiholes affects the spectrum of the edge: consider  a
sphere in a RR state with $N_e$ electrons that are all clustered
in clusters of $k$ electrons. Assume the sphere has $n$
quasiparticles and $n$ quasiholes, which are localized by
impurities away from one another. The ground state is then
degenerate, with the degeneracy being exponential in $n$ (for
large $n$). Topologically, each quasiparticle/quasihole is
equivalent to a puncture in the sphere. Now imagine bringing the
$n$ quasiparticles close together, i.e., fusing together $n$ of
the holes pierced in the sphere. The proximity of the
quasiparticles to one another lifts the degeneracy, and the way it
is lifted is determined by the state to which the $n$
quasiparticles fuse. This system is topologically equivalent  to
the system we are interested in, a disk that has $n$ quasiholes
localized in its bulk.

The energies appearing on both sides of Eq.~(\ref{coulomb-peaks})
belong to the spectrum of the edge theory. The main challenge in
obtaining the spectrum for non-Abelian states is to construct the
part of it that follows from the addition of a parafermion theory to
that of the chiral boson contributing the charging energy
Eq.~(\ref{eq:charging}). In the next section, we describe briefly
and most generally what is the relation between parafermionic field
theories and the quantum Hall effect, and who are the main players
in these theories. When discussing a particular quantum Hall state,
either one that belongs to the RR series or the spin singlet state,
we will first specialize the discussion to the parafermionic CFT
relevant to it and review its properties. We will single out the
parafermion used to construct the electron operator, and explain how
the construction of the Hilbert space is done and the spectrum is
found. Dwelling on the details of the parafermionic CFT is also
crucial in order to understand the calculation of the interference
term of Eq.~(\ref{eq:sigma xx}), as some of the phases of the two
partial waves interfering are contributed by fields in this
parafermionic theory.

\section{The CFT description of quantum Hall states}
\label{sec:CFT intro}

Conformal field theories (CFTs) are used in several different
contexts regarding quantum Hall systems. Trial many-body
wavefunctions for single Landau level fractional quantum Hall states
are created from CFTs (in $D=2+0$ dimensions) as correlators of
fields representing different particles in the system
\cite{MR,ReadRezayi,ardonne-2007-322,JainCFTforCF}. Starting from
chiral boson theories, one constructs Abelian fractional quantum
Hall wavefunctions such as those in the Laughlin series. Including
parafermionic field theories leads to wavefunctions for non-Abelian
quantum Hall states. While explicit closed form expressions are
available for RR states for general $k$ (see
Refs.~(\onlinecite{ardonne-2007-322,Cappelli,ReadRezayi})), not all
physical properties are easily explored on the basis of these
wavefunctions alone. Excitations at the edge of a quantum Hall
droplet are described by the same CFTs, now viewed in $D=1+1$
dimensions.

A theoretical construction of the paired quantum Hall state at
filling factor $1/2$ was originally proposed in the context of
CFT, and is easily understood in that language \cite{MR}. The
theoretical construction of a wavefunction for this state involves
the $\mathbb{Z}_2$ theory, also known as the Ising CFT. The use of
this CFT yields the Pffafian wavefunction, implying that the
gapped state of the bulk is similar to that of a $p_x+ip_y$
superconductor in the weak pairing phase\cite{ReadGreen}.
Therefore, this state is thought of as a superconductor, a
condensate of Cooper-pairs of composite-fermions.

The field taken from the $\mathbb{Z}_2$ theory and used to define
the electron operator for the $1/2$ state is a neutral fermion,
and in fact represents the composite fermion. For members of the
RR series with a higher value of $k$, this composite fermion is
replaced by a composite anyon, and the operator taken from the
$\mathbb{Z}_k$ theory is called a parafermion (which no longer
obeys fermionic statistics). In that case, the gapped state of the
bulk is thought of as a condensation of clusters of such anyons.
The size of the clusters is determined by the minimal number of
such parafermions which may be combined into an effective boson,
which then Bose-condense. For the $\mathbb{Z}_k$ theory this
number is $k$. The equivalent statement in the language of CFT is
that a parafermionic CFT at level $k$ can be used to describe a
state in which $k$ electrons form a cluster. This is because the
fusion rules of this parafermion theory always imply that $k$
copies of the fundamental field used to define the electron
operator fuse to the identity operator.

Since they were first theoretically constructed using parafermion
CFTs, these same CFTs remained the main available theoretical tool
to explore various properties of quantum Hall clustered states
other than the Moore-Read state. While the Moore-Read state may be
conveniently explored using the theory of superconductivity, no
analogous description exists for other clustered states.

A general parafermionic theory\cite{Gepner:1987sm} is a coset model
$G_k/U(1)^{\ell}$, where $G_k$ is a simple affine Lie algebra of
rank $\ell$. The fields are labeled $\Phi^{\Lambda}_{\lambda}$,
where both $\Lambda$ and $\lambda$ are weights of the simple Lie
algebra $G$. These fields are subject to some restrictions and
identifications, and we will specify them when dealing with a
specific parafermion theory below. The conformal dimension of these
fields is given by
\begin{equation}
h_\lambda^\Lambda=\frac{\Lambda\cdot(\Lambda+2\rho)}{2(k+g)}
                  -\frac{\lambda\cdot\lambda}{2k}+n^\Lambda_\lambda
\end{equation}
where $2\rho$ is the sum of the positive roots of the Lie algebra
$G$, and the scalar product $(\cdot)$ is with respect to the
quadratic form matrix (for more details see
Ref.~(\onlinecite{CFTbook})). The integer $n^\Lambda_\lambda$ is
equal to the grade of the representation of the current algebra in
which $\Phi_\lambda^\Lambda$ appears\cite{Gepner:1987sm}.

The parafermion field is denoted by $\psi_\alpha$, and belongs to
the set of fields $\Phi_\alpha^0$, where $\alpha$ is a root, and
the vector $0$ is the vacuum representation. The OPE of a
parafermion and a field $\Phi^{\Lambda}_{\lambda}$ is given by
\begin{equation}
\psi_\alpha(z) \Phi^\Lambda_\lambda(0)=\sum_{m=-\infty}^{\infty}
z^{-\alpha\cdot\lambda/k-1-m}[A^{\alpha,\lambda}_m\Phi^\Lambda_{\lambda}](0)
\end{equation}
Note that the product of roots $\alpha\cdot\lambda$ is with respect
to the quadratic form matrix. The modes $A^\alpha_m$ obey
generalized commutation relations.

\section{Read-Rezayi states and $\mathbb{Z}_k$ parafermions}\label{sec:RR and
parafermions} Consider a 2DEG experiencing one of the plateaus
belonging to the RR series. When the magnetic field is varied by
one flux quantum, $k$ quasiholes appear, hence the flux associated
with a single quasihole is $\frac{\phi_0}{k}$. Combined with the
fact that the filling factor is $k/(k+2)$, this implies that
quasiholes for RR states have charge $\frac{e}{k+2}$ .In order to
fully describe the edge of the RR state, one must add a second
field theory to that of the chiral boson, a CFT known as
$\mathbb{Z}_k$ parafermions \cite{ZF,Gepner:1986hr}. In this
theory the algebra $G_k$ is $SU(2)_k$ of rank $\ell=1$, and
therefore the theory is equivalent to the coset model
$SU(2)_k/U(1)$.

The fields in the theory are labeled by two quantum numbers
$\Phi^l_m$, with $l\in \left\{ 0,1,\ldots ,k\right\}$. The integer
$m$ is known as the $\mathbb{Z}_k$ charge of the field $\Phi _{m
}^{l }$ and is defined modulo $2k$. The fields are subjected to
the following identifications $\Phi _{m }^{l }=\Phi _{m
+2k}^{l}=\Phi _{m -k}^{k-l }$ and $l+m = 0 \bmod 2 $.

The creation operators of both electrons and quasiholes in the
bulk are products of two factors. The first is a vertex operator,
$e^{i\alpha\varphi(z)}$, that accounts for the flux and the charge
associated with the electron ($\alpha=\sqrt{(k+2)/k}$), and with
the quasihole ($\alpha=1/\sqrt{k(k+2)}$) (this vertex operator is
sufficient for the description of Abelian states\cite{wen}). The
second factor is one of the fields in the parafermionic theory;
the electron creation operator is given by $\psi_1
e^{i\sqrt\frac{k+2}{k}{\varphi}}$, where $\psi_1=\Phi_2^0$ is one
of the parafermionic currents, while the quasihole creation
operator is $\sigma_1 e^{i\sqrt\frac{1}{k(k+2)}{\varphi}}$, where
$\sigma_1$ is one of the parafermionic primary fields (see below).

The parafermion $\psi_1$ has the following operator product
expansion with a field of $\mathbb{Z}_k$ charge $q$
\begin{equation}
\psi_1(z)\Phi_q(0)=\sum_{p=-\infty}^{\infty}z^{-p-1-q/k}A_{(1+q)/k+p}\Phi_q(0).
\end{equation}
The fields $A_{(1+q)/k+p}\Phi_q(0)$ have $\mathbb{Z}_k$ charge
$q+2$. Similarly, the mode expansion of the parafermion
$\psi_1^{\dagger}=\psi_{k-1}$ is given by
\begin{equation}
\psi^\dagger_1(z)\Phi_q(0)=\sum_{p=-\infty}^{\infty}z^{-p-1-q/k}A^{\dagger}_{(1-q)/k+p}\Phi_q(0).
\end{equation} and the field $A^\dagger_{(1-q)/k+p}\Phi_q(0)$ has $\mathbb{Z}_k$ charge
$q-2$.

The different modes of the field $\psi_1$,
\begin{equation}
A_{(1+q)/k+p}=\frac{1}{2\pi i}\oint dz\psi_1(z)z^{q/k+p},
\end{equation}
obey generalized commutation relations, that may be found by
evaluating the integral
\begin{eqnarray}\nonumber
\frac{1}{(2\pi i)^2}\left[\oint_0 dz,\oint_0 dw\right] \psi_1(z)
z^{q/k+\mu} \, \psi_1(w)w^{q/k+\nu}(z-w)^{2/k},
\end{eqnarray}
where $\mu$ and $\nu$ are integers. These commutation relations
take the form
\begin{eqnarray}\label{eq:generalized commutators}\nonumber
\sum_{\ell\geq 0}c^{(\ell)}&&\left[
A_{(3+q)/k+\mu-\ell}A_{(1+q)/k+\nu+\ell}\right.\\
&& \left.-A_{(3+q)/k+\nu-\ell}A_{(1+q)/k+\mu+\ell}\right]=0
\end{eqnarray}
(when acting on a field with charge $q$), where
$c^{(\ell)}=\Gamma(\ell-2/k)/\ell !\Gamma(-2/k)$. The same method
can be used to obtain expressions for the generalized commutation
relation between the different modes $A^\dagger$ of
$\psi_1^\dagger$, and between the $A^\dagger$s and
$A$s\cite{ZF,Gepner:1986hr}.

The parafermionic primary fields $\Phi_l^l$ are defined by the
condition
\begin{equation}
A_{(1+\ell)/k+p}\Phi_l^l=A^\dagger_{(1-\ell)/k+p+1}\Phi_l^l=0\quad\text{for}\
p\geq 0
\end{equation}
The primary fields are usually denoted $\sigma_l$ and referred to
as spin fields. Their conformal dimension is given by
\begin{equation}\label{eq:dimension of sigma}
h_l=\frac{l(k-l)}{2k(k+2)}.
\end{equation}
Each of these fields generates a series of fields $\Phi_m^l$ by
applications of the modes $A$ of the parafermion $\psi_1$, and the
modes $A^{\dagger}$ of the parafermion $\psi_1^{\dagger}$. The
conformal dimension of the field $\Phi _{m}^{l}$ is given by
\begin{eqnarray}\nonumber
&&h^l_m=h_l+\frac{(l-m)(l+m)}{4k}, \ \text{for}\ -l<m\leq l\quad \\
&&h^l_m=h_l+\frac{(m-l)(2k-l-m)}{4k}, \ \text{for}\ l<m\leq
2k-l\quad\quad \ .\label{eq:conformal dimensions zk}
\end{eqnarray}

The fusion rules for the parafermionic CFT are
\cite{ZF,Gepner:1986hr}
\begin{equation}\label{eq:fusion}
\Phi _{m _{\alpha}}^{l _{\alpha}}\times \Phi _{m _{\beta}}^{l
_{\beta}}=\sum\limits_{l_\gamma =\left| l_{\alpha}-l _{\beta}\right|
}^{\min \left\{l _{\alpha}+l_{\beta},2k-l _{\alpha}-l
_{\beta}\right\} }\Phi _{m _{\alpha}+m _{\beta}}^{l_\gamma }.
\end{equation}
The operator product expansion (OPE) is given by
\begin{eqnarray}
\Phi _{m _{\alpha}}^{l _{\alpha}}(z) \Phi _{m _{\beta}}^{l
_{\beta}}(w)=\sum_{l_\gamma}C_{\alpha \beta \gamma}(z-w)^{\Delta h}\
\Phi _{m _{\alpha}+m _{\beta}}^{l_\gamma}(w) \label{ope}
\end{eqnarray}
where the fields appearing on the right hand side are determined
by Eq.~(\ref{eq:fusion}), $C_{\alpha \beta \gamma}$'s are
constants, and $\Delta h=h_{m _{\alpha}+m _{\beta}}^{l_\gamma
}-h_{m _{\beta}}^{l _{\beta}}- h_{m _{\alpha}}^{l _{\alpha}}$. As
a consequence of that relation, when a field $\Phi _{m
_{\alpha}}^{l _{\alpha}}$ goes around a field $\Phi _{m
_{\beta}}^{l _{\beta}}$ and their fusion is to a field $\Phi _{m
_{\alpha}+m _{\beta}}^{l_\gamma }$, the phase generated is
$2\pi\Delta h$.

\subsection{Coulomb blockade for Read-Rezayi states}
\label{subsec:coulomb blockade RR}

To determine the parafermionic part of the ground state energy for
a dot with electrons on the edge, we need to construct a basis of
states for the parafermion theory and extract from it the
lowest-lying energy state with $0\leq j<k$ parafermions. Such a
basis was constructed using the modes of the fundamental
parafermion $\psi_1$ in Ref.~(\onlinecite{mathieu2}), see also
Ref.~(\onlinecite{schoutens,bouwknegt}), and used in
Ref.~(\onlinecite{ilan}). A general state with $j$ parafermions of
the type $\psi_1$ is of the form
\begin{equation}\label{eq:state}
A_{-p_{j}+(2j+q-1)/k}A_{-p_{j-1}+(2j+q-3)/k}\cdot\cdot\cdot
A_{-p_1+(1+q)/k}\ket{\sigma_q},
\end{equation}
where the $p_i$ are integers, and $\ket{\sigma_0}=\ket{0}$. This
state is an eigenstate of the Hamiltonian with energy
\begin{eqnarray}\nonumber
E(j,q)&&=h_q - \sum_{i=0}^{j-1}\frac{q+1+2i}{k}
       +\sum_{i=1}^{j}p_i\\&&=h_q-\frac{j(j+q)}{k}+\sum_{i=1}^{j}p_i
\end{eqnarray}
in units of $\frac{2\pi v_n}{L}$, where $v_n$ is the velocity of the
neutral modes as they propagate along the edge. Since we are looking
for the lowest-energy state, the integers $p_i$ are chosen such that
the state has the lowest possible energy, and is not a null state.
The conditions are that $p_i=1$ for $i \leq k-q$ and $p_i=2$ for $i
> k-q$. In order to obtain these conditions one has to use the generalized commutation relations given
in equation (\ref{eq:generalized commutators}), see
Ref.~(\onlinecite{mathieu2}) for details. The lowest energy for a
given value of $j$ and $q$ is therefore given by
\begin{eqnarray}\label{eq:edge energy}
E_\psi(j,q)=h_q+\left\{
    \begin{array}{ccc}
    \frac{j(k-j-q)}{k} & \qquad {\rm for} & j\leq  k-q
    \\ \\
    \frac{(k-j)(j+q-k)}{k} & \qquad {\rm for} & j>  k-q\\
    \end{array}
    \right.
\end{eqnarray}
It is easy to see that for a particular value of $q$ and $j$, the
energy is simply the conformal dimension of the field which is the
result of the fusion product of $\sigma_q$ and $j$ copies of the
parafermion $\psi_1$. Also, since
\begin{equation}
\Phi^q_{q+2j}=A_{-1+(2j+q-1)/k}A_{-1+(2j+q-3)/k}\cdot\cdot\cdot
A_{-1+(1+q)/k}\Phi_q^q,
\end{equation}
for $j=k-q$ the energy $E_\psi(k-q,q)=h_q$ , i.e, the same as the
conformal dimension of the highest weight state. This is because
$\Phi^q_{q+2(k-q)}=\Phi^{k-q}_{k-q}=\sigma_{k-q}$, and the conformal
dimension of the spin field, Eq.~(\ref{eq:dimension of sigma}), is
invariant under the substitution $l\rightarrow k-l$.

We now turn to a description of the experimental procedure.
First, we characterize the state of the dot containing a quantum
Hall droplet at the beginning of the experiment, when the magnetic
field is set to some constant value, the number of electrons has
some quantized value, the area of the dot is fixed and the dot is
relaxed to its ground state.

For the edge of a quantum Hall system, the charge $q$ of the
highest weight state of the edge theory is determined by the
number of quasiholes in the bulk and the total number of electrons
in the dot, as we show below. Whether or not it stays fixed
throughout the experiment (as the area is varied), depends on the
rate at which the area is varied. This issue and its influence on
the outcome of the experiment will be discussed at later stage.

Let us start with a dot where the number of electrons is divisible
by $k$. By setting the magnetic field to a certain value for which
the filling factor is a bit less than $k/(k+2)$, quasiholes may be
introduced into the bulk, while their counterparts of opposite
charge and topological charge are introduced onto the edge. The
contribution of the charged part of the edge quasiparticles
influences the boundary conditions of the chiral boson
theory\cite{ilan}. The parafermionic part of the edge will be taken
from the fusion product of $n$ copies of the field $\sigma_{k-1}$,
the spin field that makes the edge quasiparticles.

The bulk quasiholes may have several fusion channels as dictated by
Eq.~(\ref{eq:fusion}). According to the fusion channel $\Phi_{bulk}$
of the bulk quasiholes, i.e, of $n$ copies of the field $\sigma_1$,
the fusion channel $\Phi_{edge}$ of the $n$ copies of $\sigma_{k-1}$
on the edge will be fixed by the requirement that
$\Phi_{bulk}\times\Phi_{edge}\sim\mathbf{1}$, and by the requirement
of energy minimization. We now explain how to determine which fusion
channel will ultimately be selected.

For a particular value of the magnetic field, the number of
localized quasiholes is fixed, and therefore the bulk has a single
quantized value of $\mathbb{Z}_k$ charge which is equal to $n \bmod
k $. This value of the $\mathbb{Z}_k$ charge may correspond to
several fusion channels of the parafermionic part of the quasihole
operators. Had the system with localized quasiholes been infinite
with no edge, all fusion channels of the bulk quasiholes would be
have been degenerate in energy. This is the essential ingredient
which causes these quasiholes to have non-Abelian
statistics\cite{MR,ReadRezayi}. The presence of the edge and the
excitations living on it will lift the degeneracy of these fusion
channels. The reason is that while all possible fusion channels of a
particular number of quasiholes have the same $\mathbb{Z}_k$ charge,
different fusion channels correspond to different highest weight
states with different initial occupation of parafermions on the
edge, and therefore have different energies.

Let us consider for clarity the example of $\mathbb{Z}_3$
parafermions, assumed to describe the quantum Hall state at filling
factor $\nu=12/5$. In this parafermionic theory there are two
parafermions, $\Phi_2^0=\psi_1$ and $ \Phi_4^0=\psi_2$, and two
parafermionic primary fields, $\Phi_1^1=\sigma_1$ and
$\Phi_2^2=\sigma_2$. We denote $\Phi_3^1=\Phi^2_0=\epsilon$. The
possible $\mathbb{Z}_k$ charge of the bulk is $0, 1$ or $2$.

Suppose that the number of quasiholes in the bulk is $3m+1$, where
$m$ is some integer. Then, the possible fusion channels of these
quasiholes are
\begin{equation}
(\sigma_1)^{3m+1}\sim\psi_2+\sigma_1 \ .
\end{equation}
Accordingly, the two fusion channels of the edge are $\psi_1$ and
$\sigma_2$. The two possible edge states in this case are therefore
\begin{eqnarray}
A_{-2/3}\ket{0},\quad\ket{\sigma_2}
\end{eqnarray}
with $E_\psi=2/3$ and $E_\sigma=1/15$ (in units of $2\pi v_n/L$)
respectively. If the initial edge state of the dot is set by energy
considerations, then the fusion channel of the bulk quasiholes will
be $\sigma_1$. It is easy to check that if there are $3m+2$
quasiholes in the bulk, the fusion channel of the bulk will be
$\sigma_2$, resulting in the lowest energy state $\ket{\sigma_1}$ on
the edge.

For a general value of $k$, the lowest energy of the dot will be
achieved when the edge has the lowest possible energy, under the
restriction that the $\mathbb{Z}_k$ charge of the edge state is
$k-\tilde{n}$ where we write $\tilde{n} \equiv [n]_k \equiv n
\bmod k$. The unique edge state fulfilling this requirement is the
highest weight state $\ket{\sigma_{k-\tilde{n}}}$. Therefore, the
requirement that the dot is totally relaxed into its ground state,
sets the fusion channel of the bulk quasiholes to be
$\Phi_{bulk}=\sigma_{\tilde{n}}$.

If the initial number of electrons in the dot, $N_e$,  is not
divisible by $k$, and in the absence of bulk quasiholes, there
will be a number $s$ of parafermion modes occupying edge states,
and the lowest energy of the dot will depend also on $s$. In the
background of a RR quantum Hall state, an additional electron is
represented by a bulk operator carrying a parafermion field
$\psi_1$. For the complete system to maintain zero topological
charge, this implies the presence of a corresponding dual field,
$\psi_1^\dagger=\psi_{k-1}$, on the edge. For a total of $N_e$
electrons, the edge $\mathbb{Z}_k $ charge will be $2[-N_e]_k$, as
can be seen by adding the $\mathbb{Z}_k $ charges of the
$\psi_{k-1}$ modes that come with every additional electron. Note
that it is always possible to construct states using only modes of
$\psi_1$, due to the relation
$\psi_1^{\dagger}=\psi_{k-1}\sim(\psi_1)^{k-1}$. For example, a
state with a single occupied mode of $\psi_1^\dagger$ is
equivalent to an edge state of the form (\ref{eq:state}) with
$j=k-1$ occupied modes of $\psi_1$. Therefore, the number of
occupied edge modes of the field $\psi_1$ will be $s=[-N_e]_k$.

As the magnetic field is slightly shifted, quasiholes are
introduced into the bulk.The lowest possible energy of the edge
will then be equal to the smallest conformal dimension of the
field which is a result of the fusion product of $s$ parafermions
of the type $\psi_1$ and $n$ copies of the field $\sigma_{k-1}$,
with $n$ being the number of localized quasiholes. This lowest
energy is then the smallest conformal dimension of a field with
charge $[2s-n]_k$.

The lowest conformal dimension of a field with a given
$\mathbb{Z}_k$ charge always belongs to the parafermionic primary
field of that charge. Therefore, the lowest energy of a dot with $n$
bulk quasiholes and $s$ parafermions on the edge is $h_{[2s-n]_k}$.
If the system is assumed to be initially relaxed into its absolute
ground state then the fusion channel of the bulk will be such that
all parafermionic fields on the edge will fuse to
$\sigma_{[2s-n]_k}$.

Again, let us consider an example from $\mathbb{Z}_3$ parafermions.
Suppose that the number of quasiholes in the bulk is $3m$, and the
number of electrons in the dot is $3j+1$ where $j$ and $m$ are
integers. The $\mathbb{Z}_k$ charge contributed to the edge due to
the presence of the extra bulk electron is $2[-N_e]_k=4$
corresponding to the two fields $\sigma_1$ and $\psi_2$. Using this
fact, we find that the two possible edge states are
\begin{eqnarray}
A_0A_{-2/3}\ket{0},\ket{\sigma_1}
\end{eqnarray}
with energy $E=2/3$ and $E=1/15$ respectively. Therefore, the
second state is energetically favorable. Indeed we see that the
lowest energy edge state is associated with the spin field of
charge $[2s-n]_k=1$. The fusion channel of the $3m$ quasiholes in
the bulk in this case will be set to $\epsilon$.

We conclude that the presence of a number of electrons in the dot
which is not divisible by $k$ at the beginning of the experiment
influences the fusion channel of the bulk quasiholes. It sets the
highest weight state on which parafermionic modes act.

We now turn to consider the course of the experiment. Once the
initial value of $q$ is set according to the initial occupation of
modes on the edge and the number of bulk quasiholes, the area is
varied. As an electron tunnels into the dot, the occupation of
parafermion modes on the edge changes, but we assume that the
electron does not couple to any bulk modes. While the initial
state of the dot was a highest weight state with charge
$q=[2s-n]_k$, and therefore had the lowest possible energy,
bringing the next electron into the dot results in the edge state
$A^{\dagger}_{(1-q)/k}\ket{\sigma_q}$ with charge $q-2$. This is
not necessarily the state of lowest possible energy for this
charge, since the lowest possible energy is the conformal
dimension of $\sigma_{q-2}$, and $A^{\dagger}_{(1-q)/k}\sigma_q$
is not necessarily equal to $\sigma_{q-2}$.

For the system to relax into the ground state after a change in
the area of the dot makes an electron tunnel into or out of the
dot, it has to change the fusion channel of the bulk quasiholes,
and the fusion channel of the $n$ copies of $\sigma_{k-1}$ on the
edge. Since before the electron tunneled the fusion of the
parafermionic fields on the edge was
\begin{equation}
(\psi_1)^s(\sigma_{k-1})^{n}\sim\sigma_{[2s-n]_k},
\end{equation}
and thus, that of the spin fields on the edge was
$$(\sigma_{k-1})^{n}\sim\Phi^{[2s-n]_k}_{[2s-n]_k-2s},$$
the lowest energy for an edge with one extra electron is obtained
when
\begin{equation}
\psi_{k-1}(\psi_1)^s(\sigma_{k-1})^{n}\sim(\psi_1)^{s-1}(\sigma_{k-1})^{n}\sim\sigma_{[2(s-1)-n]_k},
\end{equation}
i.e, the fusion channel of the spin fields on the edge should be
$$(\sigma_{k-1})^{n}\sim\Phi^{[2(s-1)-n]_k}_{[2(s-1)-n]_k+2(s-1)}.$$
Here, we assumed that the relaxation of the edge has to take place
only through a change in the fusion channel of the quasiholes.
This assumption amounts to assuming that no excitation with
non-trivial $\mathbb{Z}_k$ charge can tunnel between the bulk and
the edge. This point will be further discussed in Sec.
~\ref{subsubsec:Blockade RR with relaxation}.

Whether the fusion channel of the spin fields on the edge (and
therefore of the quasiholes in the bulk) changes every time the
change in the dot's area leads to a change in the number of
electrons is a question of time scales. We now turn to examine the
two limits, that of fast variation of the area of the dot where we
do not allow the edge to relax by changing the fusion channel of
the quasiholes, and that of a slow variation of the area.

\subsubsection{Zero bulk-edge relexation}

If the area of the dot is varied fast with respect to the time
scale dictated by this relaxation mechanism, then the fusion
channel of the bulk quasiholes will remain fixed throughout the
experiment.

\begin{figure}
\includegraphics[width=0.5\textwidth,angle=-90]{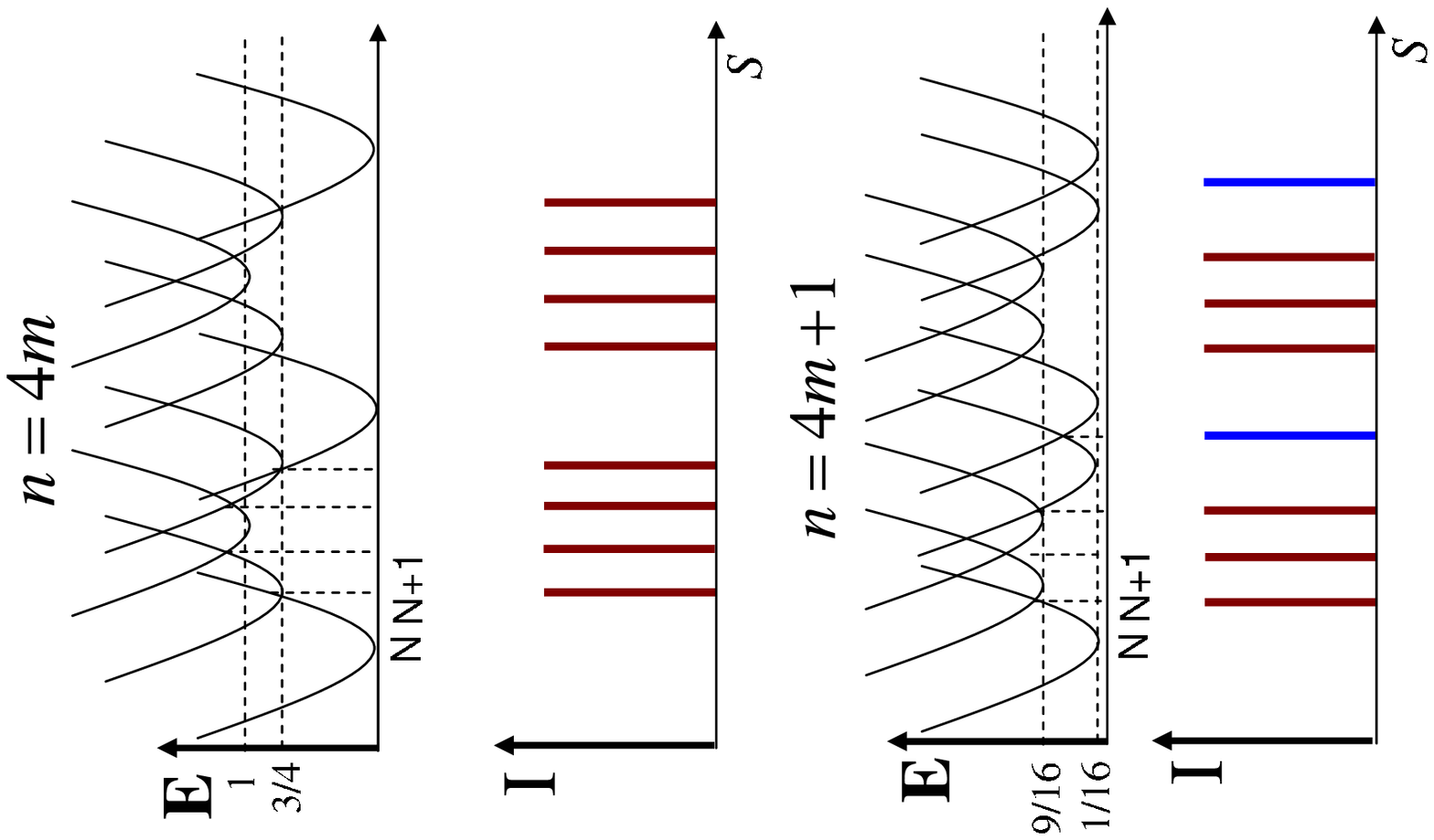}\vspace{1cm}
\includegraphics[width=0.5\textwidth,angle=-90]{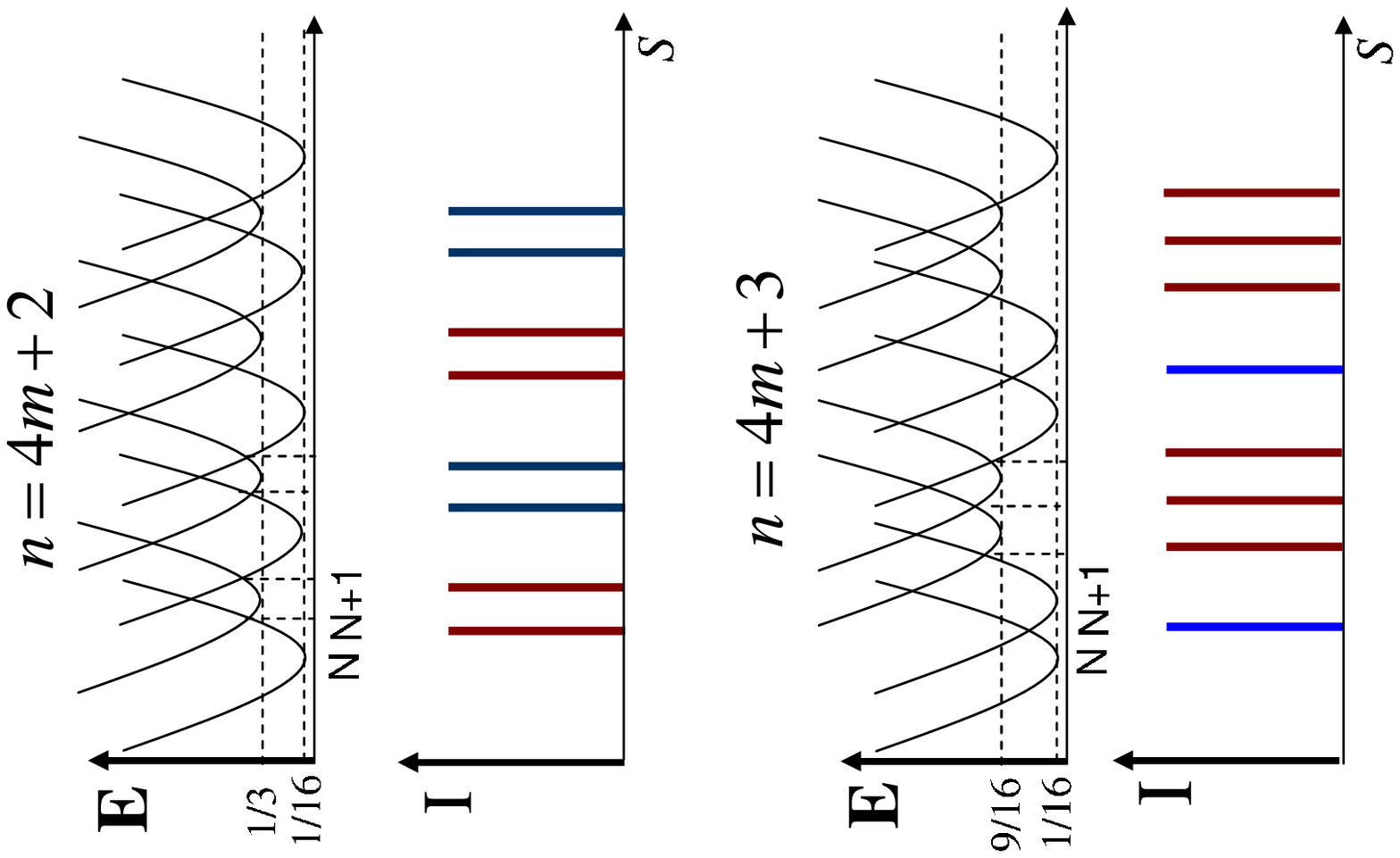}\vspace{1.7cm}
\caption{Schematic picture of the Coulomb blockade peaks for the
$k=4$ RR state (we chose $s=0$ for convenience). The total energy of
the dot, $E_c+E_\psi$, is plotted for every value of $n$ and $N_e$.
The parabolic shape of the energy curve is due to the form of the
charging energy, and the shift of the bottom of these parabolas is
given by $E_\psi$. Coulomb blockade peaks appear when two
neighboring parabolas intersect.
 } \label{fig:parabolas}
\end{figure}

In this case, as the area of the dot is varied, Coulomb blockade
peaks are observed when Eq.~(\ref{coulomb-peaks}) with
$E(N_e,S,B)=E_c(N_e,S)+E_\psi(N_e,q)$ and $q=[2s-n]_k$ is
satisfied. The spacings between Coulomb blockade peaks is easily
calculated using Eq.~(\ref{coulomb-peaks}) which assumes the form
\begin{equation}
\Delta S=\frac{e}{n_0}+\frac{eL \nu}{2 n_0 \pi
v_c}\left[E_\psi(N_e+2)-2E_\psi(N_e+1)+E_\psi(N_e)\right].
\label{area}
\end{equation}
The pattern of Coulomb blockade peaks may be described as follows:
If $q\neq 0$, the peaks bunch into alternating groups of $q$ and
$k-q$ peaks. The spacing that separates peaks within a group is
again given by
\begin{equation}\label{eq:spacing1nis}
\Delta
S_1=\frac{e}{n_0}\left(1-\nu\frac{v_n}{v_c}\frac{2}{k}\right),
\end{equation}
while the spacing that separates two consecutive groups is
\begin{equation}\label{eq:spacing2nis}
\Delta
S_2=\frac{e}{n_0}\left(1+\nu\frac{v_n}{v_c}
\left(1-\frac{2}{k}\right)\right).
\end{equation}
If $q=0$, the peaks bunch into groups of $k$. The spacing that
separates peaks within the group is again given by
(\ref{eq:spacing1nis}), while the spacing between two consecutive
groups is
\begin{equation}\label{eq:spacing2nisZero}
\Delta
S_2=\frac{e}{n_0}\left(1+\nu\frac{v_n}{v_c}\left(2-\frac{2}{k}\right)\right).\end{equation}
The periodicity of the peaks is always $k$. However, if $k$ is
even, the peak structure may also be periodic with a period $k/2$,
provided that $q=k/2$. For concreteness, the pattern is
schematically sketched in Fig.~\ref{fig:parabolas} for $k=4$.

\subsubsection{Coulomb blockade in the presence of bulk-edge relaxation}
\label{subsubsec:Blockade RR with relaxation}

If the area of the dot is varied slowly with respect to the
relaxation rate, then each electron has time to relax onto the
lowest energy state by changing the fusion channel of the edge
quasiparticles and the bulk quasiholes.

Each electron entering the dot will occupy the first mode of
$\psi_{k-1}$ operating on a highest weight state,
$\ket{\sigma_{\ell}}$, which is dictated by the number of quasiholes
in the bulk and the number of electrons $N_e$ that were inside the
dot before the tunneling of the electron took place. Therefore the
energy (in units of $2\pi v_n/L$) of the incoming electron is
\begin{equation}\label{eq:energy before decay}
h^{\ell}_{\ell-2}=h_\ell+ \frac{\ell-1}{k}.
\end{equation}
The above expression is simply the conformal dimension of the
field which is the result of the fusion product of the incoming
parafermion field, $\psi_{k-1}$, and the spin field $\sigma_\ell$
acting on the vacuum state (note that (\ref{eq:energy before
decay}) is the correct expression for $\ell>1$, otherwise the
restrictions in (\ref{eq:conformal dimensions zk}) should be taken
into account). Now, by allowing the fusion channel of the bulk
quasiholes to change, the energy of the edge state may be relaxed.
Since the current $\mathbb{Z}_k$ charge of the edge is $\ell-2$,
the fusion channel of the bulk quasihole is expected to change
such that the energy of the edge is $h_{\ell-2}$ (this means that
the fusion product of the fields on the edge changes to
$\sigma_{\ell-2}$).

In order to calculate the spacings between Coulomb blockade peaks
in this case, we use the form of equation (\ref{area}), with a
slight modification:
\begin{eqnarray}\nonumber
\Delta S=\frac{e}{n_0}+\frac{eL \nu}{2 n_0 \pi
v_c}&&\left[E_\psi(N_e+2)-E_\psi(N_e+1)\right.\\&&\left.-E'_\psi(N_e+1)+E'_\psi(N_e)\right],
\label{areaDecay}
\end{eqnarray}
where $E_\psi(N_e)$ is the energy of the edge state of a dot
containing $N_e$ electrons before it had a chance to decay, and
$E'_\psi(N_e)$ is the energy of the edge after the decay. For a
particular value of $\ell>2$, these energies are given by
\begin{eqnarray}\nonumber
&&E_\psi(N_e+2)=\frac{2\pi v_n}{L}h^{\ell-2}_{\ell-4},\ E_\psi(N_e+1)=\frac{2\pi v_n}{L}h^{\ell}_{\ell-2},\\\
&&E'_\psi(N_e+1)=\frac{2\pi v_n}{L}h^{\ell-2}_{\ell-2},\
E'_\psi(N_e)=\frac{2\pi v_n}{L}h^{\ell}_{\ell},
\end{eqnarray}
therefore the area spacing between two peaks is given again by
equation (\ref{eq:spacing1nis}). However, if the initial charge of
the edge was $\ell=1,2$, the spacing is given by equation
(\ref{eq:spacing2nis}).

The pattern of peaks we predict will be as follows. Bunching of
peaks will still be observed for all the RR states with $k\geq3$.
However, the dependence of the pattern on the number of quasiholes
in the bulk vanishes. The periodicity of this structure depends on
$k$: if $k$ is even, the periodicity is $k/2$ and the pattern is
of groups of $k/2$ peaks, while for odd $k$ the periodicity is
$k$, and the pattern is of alternating groups of $(k-1)/2$ and
$(k+1)/2$ peaks. \cite{foot1}

For the case $k=2$, we find that when inelastic bulk-edge relaxation
is allowed, the Coulomb blockade pattern will be the same as for
Abelian fractional quantum Hall states, with a constant spacing
between peaks. This means that for slow variation of the area of the
dot, the even odd effect predicted in
Ref.~(\onlinecite{SternHalperin}) is smeared out. This smearing is a
consequence of the parafermionic edge mode (which for $k = 2$ is a
Majorana fermion mode) staying unpopulated throughout the
experiment. Whenever the addition of an electron into the dot
attempts to populate this mode by a fermion, the bulk-edge
relaxation mechanism makes the fermion tunnel inelastically into the
bulk, where its presence does not involve any energy cost. It is
important to stress, as we discuss in greater detail in the summary,
that this smearing takes place only when the bulk-edge coupling
induces inelastic relaxation, and the energy of the tunneling
electron is dissipated away from the electronic system.

For other RR states, a change in the fusion channel of the bulk
quasiholes, and as a result, also of the fields on the edge, can be
understood in terms of tunneling of neutral particles as well. In
this context, neutral means having no $\mathbb{Z}_k$ charge. The
reason we allow for the bulk and the edge to exchange only particles
with zero $\mathbb{Z}_k$ charge in order to relax the energy of the
edge, is that fields that carry non trivial $\mathbb{Z}_k$ charge
are always accompanied by a vertex operator that carries real
electric charge. This physical assumption is required in order to
keep the electron operator single-valued with respect to all
possible excitation in the system. An electric charge is not allowed
to tunnel freely into the bulk due to charging energy
considerations.

\begin{figure}
\includegraphics[width=0.35\textwidth,angle=-90]{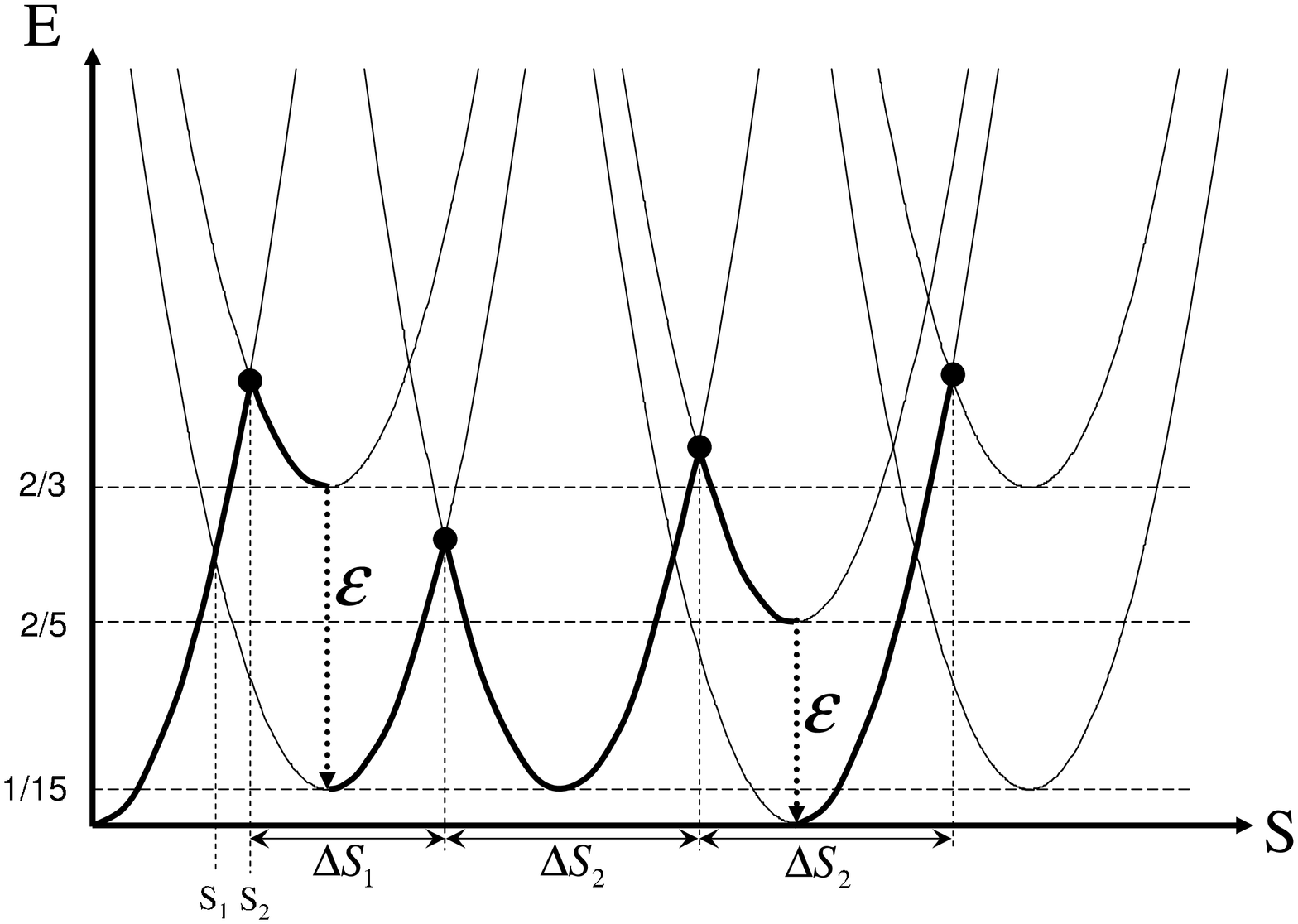}
\caption{Schematic description of bulk - edge relaxation for the
$k=3$ RR state. Every parabola on this graph represents the charging
energy as a function of the area of the dot for a given number of
electrons in the dot. The shift at the bottom of these parabolas is
the part of the energy of the edge contributed by the parafermions.
The initial state of the edge in this example is the vacuum state.
Crossing points between neighboring parabolas that correspond to
electron-tunneling events are marked by a black
dot. The thick black curve represents the energy of the edge for any
given value of the area. Note that two parabolas centered around the
same point correspond to the same number of electrons in the dot,
but a different fusion channel of the bulk quasiholes. A decay
between two parabolas is an exchange of an $\varepsilon$ field
between the edge and the bulk. We do not allow decay processes that
require the tunneling and the decay to occur simultaneously, as in
the region between the crossing points marked above $S_1$ and $S_2$.
} \label{fig:relaxationZ3}
\end{figure}

Some more intuition on this relaxation process may be gained from
Fig.~\ref{fig:relaxationZ3}, where we plot the energy as a function
of the area. This graph, similarly to the graph presented in
Fig.~\ref{fig:parabolas}, shows a set of parabolas representing the
energy of the dot as a function of its area. However, this time,
different parabolas for the same value of $N_e$, corresponding to
two different fusion channels of the parafermion fields on the edge,
may participate in the process. Switching between two such parabolas
is done by exchanging a neutral particle between the edge and the
bulk. Note that only the crossing points marked by a black dot are
those that indicate the location of a Coulomb peak. Other crossing
points, such as the one denoted by $S_1$, correspond to higher order
events we neglect, since both tunneling of an electron into the dot
and relaxation take place simultaneously. Naturally, these processes
are expected to occur on a longer time scale than relaxation alone,
and are therefore at this point excluded from the discussion.

\section{non-Abelian spin singlet states}\label{sec:AS states}

The non-Abelian spin singlet states (NASS) at $\nu=\frac{2k}{2k+3}$
are close analogues of the clustered (parafermion) states of Read
and Rezayi. The main difference is the role of the electron spin
degree of freedom: while the RR states describe spin-polarized
electrons, the spin-singlet states describe unpolarized electrons
which make up a state that is a singlet under the spin $SU(2)$
symmetry. The quasiholes of smallest fractional charge
($q=\frac{1}{2k+3}$) are spin-1/2 particles. For $k>1$ they carry
non-Abelian statistics. In the same way that the RR states
generalize the $\nu=1/3$ Laughlin state, the $\nu=\frac{2k}{2k+3}$
spin-singlet states can be thought of as generalizations of the
Abelian Halperin state at $\nu=2/5$.

Clustering of electrons in spin-singlet states is rather different
and a bit more complicated than in the RR states. The RR state at
$\nu=k/(k+2)$ can be thought of as a quantum fluid made of clusters
of $k$ (spin polarized) electrons. The formation of such clusters is
mirrored quite clearly in the relation $(\psi_1)^k\sim\mathbf{1}$
satisfied by the fundamental parafermion $\psi_1$. The spin singlet
states are associated with more complicated parafermion theories,
formally denoted as $SU(3)_k/U(1)^2$. There are now two fundamental
parafermions $\psi_1$ (associated to spin-up electrons) and $\psi_2$
(spin-down). The fusion rule $(\psi_1 \psi_2)^k\sim\mathbf{1}$
indicates that the smallest cluster with total spin zero is now
composed of $2k$ electrons.

We will focus on the simplest example, which is the state with
$k=2$ with filling fraction $\nu=4/7$. The $SU(3)_2/U(1)^2$
parafermion theory has central charge $c=6/5$. The $k=2$ state is
made up of clusters of four electrons rather than two, each
cluster having total spin zero. The CFT describing this state
employs fields that are products of free boson vertex operators
and parafermionic fields. There are now two fundamental bosons:
$\varphi_c$ corresponding to charge and $\varphi_s$ giving the
spin degrees of freedom. The parafermionic fields are again the
source of the non-Abelian statistics of the quasiparticles.

For the $SU(3)_2/U(1)^2$ parafermionic theory, the parafermionic
charge is a two component vector, and each field in the theory is
labeled by two vectors,
$\Phi^{\Lambda}_{\lambda}=\Phi^{(\Lambda_1,\Lambda_2)}_{(\lambda_1,\lambda_2)}$.
The identification rules for the fields in this case
are\cite{ardonne-2007-322}
\begin{eqnarray}
\begin{array}{ll}
  \Phi^{(\Lambda_1,\Lambda_2)}_{(\lambda_1,\lambda_2)}\equiv\Phi^{(\Lambda_1,\Lambda_2)}_{(\lambda_1+4,\lambda_2-2)} \ ,
& \Phi^{(\Lambda_1,\Lambda_2)}_{(\lambda_1,\lambda_2)}\equiv\Phi^{(\Lambda_1,\Lambda_2)}_{(\lambda_1-2,\lambda_2+4)} \ ,
\\[2mm]
  \Phi^{(\Lambda_1,\Lambda_2)}_{(\lambda_1,\lambda_2)}\equiv\Phi^{(2-\Lambda_1-\Lambda_2,\Lambda_2)}_{(\lambda_1+2,\lambda_2)} \ ,
& \Phi^{(\Lambda_1,\Lambda_2)}_{(\lambda_1,\lambda_2)}\equiv\Phi^{(\Lambda_1,2-\Lambda_1-\Lambda_2)}_{(\lambda_1,\lambda_2+2)} \ . \\
\end{array}
\end{eqnarray}
Due to these identifications, the parafermion theory has eight
fields, which we label in accordance with the notation of
Ref.~(\onlinecite{ardonne-2007-322})
\begin{eqnarray}\label{eq:SU(3)2 pararefmions}
\begin{array}{cccc}
  \mathbf{1}=\Phi^{(0,0)}_{(0,0)}  ,
  &\psi_{1}=\Phi^{(0,0)}_{(2,-1)}  ,
  &\psi_{2}=\Phi^{(0,0)}_{(-1,2)}  ,
  &\psi_{12}=\Phi^{(0,0)}_{(1,1)}  , \\
  \ &\ \\
  \sigma_{\da}=\Phi^{(1,1)}_{(2,-1)}  ,
  &\sigma_\ua =\Phi^{(1,1)}_{(-1,2)}  ,
  &\sigma_3 =\Phi^{(1,1)}_{(1,1)}  ,
  &\rho=\Phi^{(1,1)}_{(0,0)}  . \\
\end{array}
\end{eqnarray}
These fields obey the fusion rules presented in table \ref{fusion
rulas su3}. Their conformal dimensions are
\begin{equation}
h_\psi=\frac{1}{2},\quad h_\sigma=\frac{1}{10},\quad h_\rho=\frac{3}{5}.
\end{equation}

We write $\vec{\varphi}=(\varphi_c,\varphi_s)$, where $\varphi_c$ and $\varphi_s$
are bosonic fields. Using the notation
\begin{equation}
\vec{\alpha}_{\uparrow}=\frac{1}{2}\left(\frac{1}{\sqrt{7}},1\right),
\
\vec{\alpha}_{\downarrow}=\frac{1}{2}\left(\frac{1}{\sqrt{7}},-1\right),
\ \vec{\alpha}_3=\left(\frac{1}{\sqrt{7}},0\right), \nonumber
\end{equation}
the three quasihole operators are
\begin{equation}
\label{eq:quasiholes}
V_{qh}^{\uparrow}=\sg_{\ua}e^{i\vec{\alpha}_{\ua}\cdot\vec{\varphi}},
\quad
V_{qh}^{\downarrow}=\sg_{\da}e^{i\vec{\alpha}_{\da}\cdot\vec{\varphi}},
\quad
V_{qh}^{0}=\sg_{3}e^{i\vec{\alpha}_{3}\cdot\vec{\varphi}},
\end{equation}
corresponding to a spin-up quasihole with charge $1/7$, a
spin-down quasihole with charge $1/7$, and a spin-less quasihole
with charge $2/7$, respectively. The operators creating the two
type of spinful electrons in the system are
\begin{equation}
V_{el}^\ua=\psi_1 e^{i(\sqrt{7/4}\varphi_c+1/2\varphi_s)} \ ,
\quad
V_{el}^\da=\psi_2 e^{i(\sqrt{7/4}\varphi_c-1/2\varphi_s)},
\end{equation}

Given this form of the electron operators, it is evident from the
fusion rules that it takes four electrons to create a cluster with
zero spin, since $\psi_1\times\psi_2\neq\mathbf{1}$.

\begin{table}
\begin{tabular}{|c|c|c|c|c|c|c|c|c|}
\hline $\times$& & $\sigma_\uparrow$ &
$\sigma_\downarrow$&$\sigma_3$ & $\rho$ & \quad $\psi_1$\quad &\quad
$\psi_2$\quad &\quad $\psi_{12}\quad $
\\ \hline\hline
$\sigma_\uparrow$& & $\mathbf{1} + \rho$ &&&&&&  \\\hline

$\sigma_\downarrow$& & $\psi_{12}+\sigma_3$ & $\mathbf{1}+\rho$
&&&&&\\\hline

$\sigma_3$& & $\psi_1 + \sigma_\downarrow$ & $\psi_2 +
\sigma_\uparrow$ & $\mathbf{1} + \rho$ &&&&\\\hline

$\rho$& & $\psi_2$ + $\sigma_\uparrow$ & $\psi_1 +
\sigma_\downarrow$ & $\psi_{12} + \sigma_3$ & $\mathbf{1}+ \rho$
&&&\\\hline

$\psi_1$& & $\sigma_3$ & $\rho$ & $\sigma_\uparrow$ &
$\sigma_\downarrow$ & $\mathbf{1}$ &&\\\hline

$\psi_2$& & $\rho$ & $\sigma_3$ & $\sigma_\downarrow$ &
$\sigma_\uparrow$ & $\psi_{12}$ & $\mathbf{1}$ &\\\hline

$\psi_{12} $& & $ \sigma_\downarrow$ & $\sigma_\uparrow$ & $\rho$ &
$\sigma_3$ & $\psi_2$ & $\psi_1$ & $\mathbf{1}$ \\\hline
\end{tabular}
\vskip 2mm \caption{Fusion rules of the parafermion and spin fields
associated to the parafermion theory SU$(3)_2/[$U(1)$]^2$ .}
\label{fusion rulas su3}
\end{table}
\subsection{Coulomb blockade regime}\label{subsec:coulomb blockade su3}

As we have mentioned above, the field theory describing the
dynamical edge modes of the spin-singlet state at $\nu=4/7$ is a
sum of three theories: two free chiral bosons, and a parafermionic
field theory. The spectrum of the edge is therefore also made of
three contributions
\begin{equation}
E=E_c+E_s+E_\psi
\end{equation}
where $E_c$ labels the contribution of the charge boson $\varphi_c$,
$E_s$ labels the contribution of the spin boson $\varphi_s$, and
$E_\psi$ is the contribution from the parafermionic theory.

The contribution to the spectrum of the edge coming from the
parafermions is again calculated by constructing the Hilbert space
of parafermionic states, the set of highest weight states on which
creation modes of the parafermions $\psi_1$ and $\psi_2$ operate.

The mode expansion of the parafermion
$\psi_\alpha\equiv\Phi^{(0,0)}_{(\alpha_1,\alpha_2)}$ is given by
\begin{equation}\label{eq:mode expansion}
\psi_\alpha=\sum_{m}z^{-m-\frac{1}{2}}\psi^{(\alpha)}_{m},
\end{equation}
where the mode indices $m$ may be integer or half-integer,
depending on boundary conditions of the parafermion set by the
field on which these parafermions act. For example, when acting on
the vacuum state, $m\in \mathbb{Z}+1/2$, since the parafermion
should obey periodic boundary conditions. When acting on any other
field, $\Phi^\Lambda_\lambda$, the boundary conditions can easily
be determined using the Operator Product Expansion between the two
fields.

The charging energy as a function of the number of electrons in the
dot may be found using the same considerations that were explained
in Sec.~\ref{sec:experimental}. It is given by Eq.~(\ref{eq:charging}).
The energy cost associated with creating non-zero spin inside the dot
will, by analogy, have the form
\begin{equation}
\label{eq:spin energy}
E_s=\frac{v_s}{4\pi}\int\left(\partial_x\varphi_s
\right)^2dx=\frac{\pi v_s}{ 4L}N_s^2
\end{equation}
with $N_s$ being the net number of unpaired spins on the edge.

We assume that the region outside the dot, the leads from which
electrons tunnel into and out of the dot, contains electrons of
both spins, and that both spins are equally available, such that
finding the lowest energy state of $j$ parafermions on the edge is
not constrained by availability of a certain type of spin. Since
the ground state of the dot is a spin singlet, breaking "spin
neutrality" must cost a certain amount of energy, and this energy
cost given by Eq.~(\ref{eq:spin energy}) will influence the
order in which electrons enter the dot.

Using the same argument we used in section \ref{subsec:coulomb
blockade RR}, we now turn to construct the lowest lying energy
states of the parafermion theory. This time, however, we must take
into account the spin of the incoming electron in order to
determine the order in which the parafermion modes are applied to
the highest weight state. While the charging energy is
unavoidable, the other two components of the energy may compete
with each other. We now turn to evaluate a particular example in
order to demonstrate the effect of this competition.

We start from the case when the bulk of the dot does not include
any quasiholes ($n=0$). The fusion rules and the requirement that
the ground state of the dot is a spin singlet imply that the
electrons cluster into groups of $4$. If the total number of
electrons in the dot is not divisible by $4$, the remainder
accumulate at the edge, occupying charge, spin and parafermionic
modes. The parafermionic state of the edge is then obtained by
applying $j$ parafermion operators to the vacuum, with $0\leq
j\leq 3$. For simplicity, we start from a situation where the
initial number of spin-up electrons, $N^0_\ua$, is equal to the
number of spin down electron, $N^0_\da$, which is even. Therefore
the highest weight state of the edge theory is $\ket{0}$, the
vacuum state, and the initial number of occupied edge modes is
zero.

The first electron to tunnel into the dot will create non-zero total
spin inside it. Since the energy cost involved, $E_s=\pi v_s/4L$,
cannot be avoided and is the same for both types of spin we choose
without loss of generality that this electron is a spin up electron.
The modes $m$ of the parafermion part of the electron operator,
$\psi_1$, will be half-integer, and therefore the state with one
occupied parafermion mode is
\begin{equation}
\psi^{(1)}_{-\frac{1}{2}}\ket{0}
\end{equation}
and the corresponding energy is $E_\psi=1/2(2\pi v_n/L)$.

We now estimate the energy cost for the tunneling of the second
electron assuming that the first one was a spin up electron
($\psi_1$). If the second electron is a spin up electron, we will
again need to pay energy for increasing the total spin of the
system. The boundary conditions on the $\psi_1$ field are periodic,
and therefore the parafermionic edge state will be
\begin{equation}\psi^{(1)}_{1/2}\psi^{(1)}_{-1/2}\ket{0},
\end{equation}
with $E_\psi=0$ and $E_s=4(\pi v_s/4L)$.

On the other hand, if the second electron is a spin down electron,
the spin of the system will be reduced back to zero. This way, the
energetic cost associated with spin polarization is avoided. The
boundary conditions on the $\psi_2$ field are antiperiodic, since
$\psi_1\cdot\psi_2\sim z^{-1/2}\psi_{12}$, and therefore the state
will be
\begin{equation}
\psi^{(2)}_{0}\psi^{(1)}_{-1/2}\ket{0},
\end{equation}
with $E_\psi=1/2(2\pi v_n/L)$, and the energy due to spin is
$E_s=0$.

Comparing the two scenarios in which two electrons tunneled into
the edge of the system, we find that in order to determine which
state has lower energy we must consider the ratio between $v_n$
and $v_s$. Bringing in a spin down electron as the second electron
is an energetically favorable process as long as $v_n/v_s<1$,
while when $v_n/v_s>1$, the lowest energy will correspond to a
state of two electrons with the same spin.

If $v_n/v_s<1$, then the state of three electrons on the edge must
correspond to
\begin{equation}
\psi^{(1)}_{0}\psi^{(2)}_{0}\psi^{(1)}_{-1/2}\ket{0},
\end{equation}
with $E_\psi=1/2(2\pi v_n/L)$ and $E_s=(\pi v_s/4L)$. Finally, the
fourth electron entering the dot will have spin down, setting the
total spin of the dot back to zero. The parafermion state of four
parafermions is identified with the vacuum state.

Since the lowest energy of a state with $j$ parafermions is
always given by the conformal dimension of the field which is the
result of the fusion product of all the parafermions, the
tunneling sequences can intuitively be described using a simple
diagram. The sequence corresponding with the case $v_n/v_s>1$, can
be represented diagrammatically as
\begin{equation}\label{eq:identity sequence 1}
\mathbf{1}\stackrel{\psi_1}\longrightarrow
\psi_1\stackrel{\psi_1}\longrightarrow
\mathbf{1}\stackrel{\psi_2}\longrightarrow\psi_2\stackrel{\psi_2}\longrightarrow
\mathbf{1}
\end{equation}
The parafermion above the arrow in the diagram is the parafermion
added to the edge state, and the fields at the tip of an arrow is
the result of the fusion product of the field at the beginning of
the arrow with that parafermion. The energies are given by
\begin{eqnarray}
&&E_0=E_c(N_e) \nonumber \\
&&E_1=E_c(N_e+1)+\frac{\pi v_n}{ L}+\frac{\pi v_s}{4L} \nonumber \\
&&E_2=E_c(N_e+2)+\frac{\pi v_s}{ L} \nonumber \\
&&E_3=E_c(N_e+3)+\frac{\pi v_n}{ L}+\frac{\pi v_s}{4L}
\end{eqnarray}

The sequences corresponding to the case $v_n/v_s<1$ are either
\begin{equation}\label{eq:identity sequence 2}
\mathbf{1}\stackrel{\psi_1}\longrightarrow
\psi_1\stackrel{\psi_2}\longrightarrow
\psi_{12}\stackrel{\psi_1}\longrightarrow\psi_2\stackrel{\psi_2}\longrightarrow
\mathbf{1},
\end{equation}
or alternatively
\begin{equation}\label{eq:identity sequence 3}
\mathbf{1}\stackrel{\psi_1}\longrightarrow
\psi_1\stackrel{\psi_2}\longrightarrow
\psi_{12}\stackrel{\psi_2}\longrightarrow\psi_1\stackrel{\psi_1}\longrightarrow
\mathbf{1}
\end{equation}
And this time the energies are given by
\begin{eqnarray}
&&E_0=E_c(N_e) \nonumber \\
&&E_1=E_c(N_e+1)+\frac{\pi v_n}{ L}+\frac{\pi v_s}{4L} \nonumber \\
&&E_2=E_c(N_e+2)+\frac{\pi v_n}{L}\nonumber \\
&&E_3=E_c(N_e+3)+\frac{\pi v_n}{ L}+\frac{\pi v_s}{4L}
\end{eqnarray} Of course, interchanging $\psi_1$ with $\psi_2$ in
these diagrams leaves the energies at each stage completely
invariant.

Using Eq.~(\ref{eq:charging}) and $(\ref{coulomb-peaks})$,
we may find the area, which appears in the expression for the
charging energy, for which transport through the dot is allowed.
Then we may calculate the area spacing between peaks. We find that
the two sequences above correspond to two different Coulomb peak
structures. The first sequence corresponds to
\begin{eqnarray}\label{eq:spacings for identity 1}
&&\Delta
S_1=\frac{e}{n_0}\left(1+\nu\frac{v_s}{4v_c}-\nu\frac{v_n}{v_c}\right)
\nonumber \\
&&\Delta
S_2=\frac{e}{n_0}\left(1- 3 \nu\frac{v_s}{4v_c}+\nu\frac{v_n}{v_c}\right)
\nonumber \\
&&\Delta
S_3=\frac{e}{n_0}\left(1+\nu\frac{v_s}{4v_c}-\nu\frac{v_n}{v_c}\right)
\nonumber \\
&&\Delta
S_4=\frac{e}{n_0}\left(1+\nu\frac{v_s}{4v_c}+\nu\frac{v_n}{v_c}\right)
\end{eqnarray}
while the second sequence corresponds to
\begin{eqnarray}
\label{eq:spacings for identity 2}
&&\Delta
S_1=\frac{e}{n_0}\left(1-\nu\frac{v_s}{4v_c}-\frac{1}{2}\nu\frac{v_n}{v_c}\right) \nonumber \\
&&\Delta
S_2=\frac{e}{n_0}\left(1+\nu\frac{v_s}{4v_c}\right) \nonumber \\
&&\Delta
S_3=\frac{e}{n_0}\left(1-\nu\frac{v_s}{4v_c}-\frac{1}{2}\nu\frac{v_n}{v_c}\right) \nonumber \\
&&\Delta
S_4=\frac{e}{n_0}\left(1+\nu\frac{v_s}{4v_c}+\nu\frac{v_n}{v_c}\right)
\end{eqnarray}
These spacings repeat themselves as the area is varied. It is
obvious that they fluctuate around the value anticipated to occur
for the Abelian states, just as in the case of the RR series.

It is instructive to compare the $k=2$ spin-singlet state to the
$k=2$ RR state (a.k.a. the Moore-Read state). While both states
can be obtained as maximal density zero-energy eigenstate of the
same `pairing' Hamiltonian \cite{StateCounting}, the periodicity
in the Coulomb blockade peak structure is different. For the
Moore-Read the maximal periodicity is 2 while for the spin singlet
state we find a periodicity of 4, in agreement with the physical
picture where the state is built up from clusters having 4
electrons each.

When $n$ is non-zero, the requirement for spin-neutrality leads to
$N_\ua+n_\ua=N_\da+n_\da$, where $n_{\ua(\da)}$ denotes the number
of spin-up (down) quasiholes and $N_{\ua(\da)}$ denotes the number
of spin-up (down) electrons . By the same considerations applied for
the RR states, the fusion channel of the bulk quasiholes directly
influences the fusion channel of the spin fields on the edge, thus
determining the highest weight state on which the modes of the
parafermions act. Also, these fusion channels at the beginning of
the experiment are fixed such that the edge has the lowest possible
energy.

Looking at the fusion rules (Table~\ref{fusion rulas su3}), we find
that the fusion channel of the bulk quasiholes will be the same as
the fusion channel of the spin fields on the edge. Having an initial
number of spin up electrons on the edge, given by $[N_\uparrow]_2$,
or  spin down electrons, given by $[N_\downarrow]_2$, will influence
both fusion channels.

Let us consider an example for concreteness. Suppose that the
possible fusion channels of quasiholes in the bulk are
$\psi_2+\sigma_\ua$. If $[N_\uparrow]_2=[N_\downarrow]_2=0$, the
requirement that the edge has the lowest possible energy fixes the
fusion channel of the edge to be $\sigma_\ua$, since it has a lower
conformal dimension. However, if this is not the case, the fusion
channel of the bulk quasiholes may be $\psi_2$, such that the edge
is in the vacuum state. This may occur when $[N_\uparrow]_2=0$ but
$[N_\downarrow]_2=1$.

In general, there are four possible highest weight states on the
edge: $\ket{0}, \ket{\sigma_\ua}, \ket{\sigma_\da}$ and
$\ket{\sigma_3}$, and it is difficult to predict how they
correspond to the exact number of quasiholes in the dot, and to
the initial number of electrons on the edge, as we have done for
the RR states. This is true mostly because there may also be
quasiholes of zero spin in the bulk which do not contribute to the
total spin, but may change the possible fusion channel of the bulk
(and therefore the one of the edge). Consequently, we study
tunneling of electrons into an edge characterized by each of these
highest weight states, knowing that initially, the total spin of
the dot was zero, and by applying the same logic we applied for
the vacuum state above. The results are summarized in
Table~\ref{tab:tunneling sequences}.

If the system is allowed to to relax into its ground state after
each tunneling event of an electron into the dot, i.e, the fusion
channel of the bulk quasiholes is allowed to change when the new
electron is added, the sequences appearing in
Table~\ref{tab:tunneling sequences} will change whenever one of
the fields in the chain is not a primary field. Noting that, we
realize that the sequences that may change are the ones starting
with the identity field, and the second sequence constructed from
the primary field $\ket{\sigma_3}$, appearing last on
Table~\ref{tab:tunneling sequences}.
When drawing the new sequences we will denote a decay to a lower
energy state (switching to a field with the same charge, but with
lower conformal dimension) by a dashed arrow.

Let us see how this generalization applies to the sequences that
formally started with the identity operator. We therefore start with
the case where the fusion channel of the bulk quasiholes is such
that the initial state of the edge is either $1$ or $\rho$, and may
switch between these two in order to minimize the energy. This
switching is again done via the exchange of a neutral particle, as
we discussed at the end of Sec.~\ref{subsec:coulomb blockade RR}. In
this case, the neutral particle is $\rho$. Before any
electron-tunneling occurs, the state of the edge will be the vacuum
state, and therefore our sequence begins with the identity field
again. It will be, for example
\begin{equation}\label{eq:decay form identity su3}
\mathbf{1}\stackrel{\psi_1}\longrightarrow
\psi_1\dashrightarrow\sigma_\da\stackrel{\psi_2}\longrightarrow
\sigma_3\stackrel{\psi_1}\longrightarrow\sigma_\ua\stackrel{\psi_2}\longrightarrow
\rho\dashrightarrow\mathbf{1}
\end{equation}
for the limit in which the spin energy is the dominant energy
($v_n/v_s<1$). The energy cost for each of the tunneling electrons
and the energy of the edge after relaxation takes place are again
easily calculated using the conformal dimension of the parafermionic
fields as well as Eqs.~(\ref{eq:charging}) and (\ref{eq:spin
energy}). Other ordering of $\psi_1$ and $\psi_2$ that minimize the
spin energy will yield the same set of energies. Plugging these into
equation (\ref{areaDecay}), we then find that the set of spacings
between peaks changes from the one appearing in
Eq.~(\ref{eq:spacings for identity 1}) to the following set
\begin{eqnarray}\label{eq:identity spacings with decay 1}
&&\Delta
S_1=\frac{e}{n_0}\left(1+\nu\frac{v_s}{4v_c}\right)\\\nonumber
&&\Delta
S_2=\frac{e}{n_0}\left(1-\nu\frac{v_s}{4v_c}+\frac{1}{2}\nu\frac{v_n}{v_c}\right)\\\nonumber
&&\Delta
S_3=\frac{e}{n_0}\left(1+\nu\frac{v_s}{4v_c}\right)\\\nonumber
&&\Delta
S_4=\frac{e}{n_0}\left(1-\nu\frac{v_s}{4v_c}-\frac{1}{2}\nu\frac{v_n}{v_c}\right)
\end{eqnarray}
which is slightly different but again corresponds to a periodicity
of $4$ for the Coulomb Blockade structure.

In the opposite limit where the energy associated with the spin
excitation is smaller, the sequence of fields will be
\begin{eqnarray}\label{eq:identity spacings with decay 2}
\mathbf{1}\stackrel{\psi_1}\longrightarrow
\psi_1\dashrightarrow&&\sigma_\da\stackrel{\psi_2}\longrightarrow
\sigma_3\stackrel{\psi_1}\longrightarrow\\&&\sg_\ua\stackrel{\psi_1}\longrightarrow\sg_3\stackrel{\psi_2}\longrightarrow\sg_\da\stackrel{\psi_2}\longrightarrow\sg_3\stackrel{\psi_1}\longrightarrow\sg_\ua\stackrel{\psi_1}\longrightarrow\sg_3....\nonumber
\end{eqnarray}
and so on. The above sequence shows that after the first electron
tunnels into the dot, the sequence turns out to be the same as if
the highest weight state of the dot was $\ket{\sg_3}$, and
therefore the spacings will be the same as for that case (see
Table~\ref{tab:tunneling sequences}).

When the fusion channel of the bulk quasiholes is such that the
initial state of the edge is $\sigma_3$, and the spin energy is the
dominant one, the sequence of fields generalizes to
\begin{equation}
\sigma_3\stackrel{\psi_1}\longrightarrow
\sigma_\ua\stackrel{\psi_2}\longrightarrow \rho\dashrightarrow
\mathbf{1}\stackrel{\psi_1}\longrightarrow\psi_1\dashrightarrow\sigma_\da\stackrel{\psi_2}\longrightarrow
\sigma_3.
\end{equation}
Note that this sequence is identical to sequence (\ref{eq:decay form
identity su3}), and following from it is the same set of spacings
given in Eq.~(\ref{eq:identity spacings with decay 1}).

In conclusion, we have found that depending on the highest weight
state of the edge, the periodicity of the Coulomb blockade peak
pattern will be either $4$ or $2$.
Also, in contrast with the result we obtained for the $k=2$ RR
state, where the time scale on which the area is varied strongly
influenced the periodicity, we find that for the state at
$\nu=4/7$, while the spacings themselves are affected by the
relaxation of the edge, the periodicity of the peak pattern
remains unaffected.

\subsection{Lowest order interference}
\label{subsec:LOI in AS states}

In this section we find the interference term of Eq.~(\ref{eq:sigma
xx}) when the bulk is at the $\nu=4/7$ NASS state. We use the CFT
input detailed at the beginning of Sec.~\ref{sec:AS states}.

Lowest order interference for the RR series of states was studied
before in Ref.~(\onlinecite{Bonderson2}), and a general expression
for the interference term was obtained. The approach of
Ref.~(\onlinecite{Bonderson2}) relied on studying the properties
of the modular S-matrix relevant for the parafermion and free
chiral boson CFTs.

For convenience, in this section we shall refer to parafermionic
fields with a fusion product that may yield more than one possible
outcome as "non-Abelian" fields (those are the three $\sigma$'s
and $\rho$), and to those who always fuse to one field as
"Abelian" (the three $\psi$'s and $\mathbf{1}$). As we will see,
there will be a crucial difference between the interference term
in the case where the fusion product of the bulk quasiholes
results in a non-Abelian or an Abelian filed. This is also true
for the RR states, as was demonstrated before in
Refs.~(\onlinecite{SternHalperin,Bonderson1,Bonderson2,Stone}).

Every one of the $n$ localized bulk quasiholes, and also the
quasiholes propagating along the edge are either spin-up,
spin-down, or spin-less, corresponding to the operators listed in
Eq.~(\ref{eq:quasiholes}). Again, we denote by $n_\ua$ the number
of localized quasiholes with spin up and by $n_\da$ the number of
localized quasiholes with spin down.  The number of localized
spin-less quasiholes is denoted $n_3$, such that
$n=n_\ua+n_\da+n_3$.

The result of the fusion product of all the quasiholes localized in
the bulk into a single field can be obtained by fusing all the
bosonic contributions into the single bosonic operator,
$e^{i\frac{1}{2}\left(\frac{n_{\ua}+n_{\da}+2n_{3}}{\sqrt{7}},
n_{\ua}-n_{\da}\right)\cdot\vec{\varphi}}$, and fusing all the spin
fields of the quasihole operators into a single parafermionic field.
The operator that is obtained represents the internal state of the
island between the two point contacts, and will ultimately determine
the form of the interference term we wish to calculate.

We assume here that the bulk quasiholes have a definite fusion
channel. Although in principle the fusion product of all the spin
fields has two possible outcomes, it was shown in a previous work
\cite{Bonderson3} that the measurement procedure collapses this
superposition onto a particular fusion channel. We do not elaborate
on the collapse process here, but rather refer the interested reader
to Ref.~(\onlinecite{Bonderson3}) for more details.

The fusion product of $n$ spin fields of different types may have
eight different results, due to the existence of eight parafermionic
fields of the $SU(3)_2/U(1)^2$ coset. Accordingly, the state of the
island may be represented by eight different operators. For a fixed
value of $n_{\ua}$, $n_{\da}$ and $n_{3}$, the state of the island
will be one of two fields, Abelian or non-Abelian according to the
fusion rules (Table~\ref{fusion rulas su3}).

We now study the interference between the two partial waves
$\ket{\psi_L}$ and $\ket{\psi_R}$ discussed in
Sec.~\ref{sec:experimental}, and examine how different internal
states of the dot affect them, hence influencing the interference
term of the backscattered current.

When the internal state of the island is represented by a
non-Abelian field, the fusion product of this field with an
incoming quasihole on the edge is a superposition of two fields.
To each of these we assign a state. We denote the two states by
$\ket{0}$ and $\ket{1}$, referring to an Abelian and non-Abelian
field correspondingly. For example, if there is only one localized
spin up quasihole in the bulk, and the edge quasihole also has
spin up, the two spin fields of the type $\sg_\ua$ fuse as
follows: $\sg_\ua\sg_\ua\sim 1+\rho$. Therefore, we say that the
system is in a superposition state
$\ket{\psi}=a_0\ket{0}+a_1\ket{1}$, with $|a_0|^2+|a_1|^2=1$. The
state $\ket{0}$ corresponds to the Abelian field $\mathbf{1}$, and
the state $\ket{1}$ corresponds to the non-Abelian field $\rho$.
Note that the two partial waves $\ket{\psi_{L,R}}$ describe the
state of the $n$ localized quasiholes and the incoming edge
quasihole.

The two partial waves can be written, using the above notation, as
follows
\begin{eqnarray}
&&\ket{\psi_L}=a_0\ket{0}+a_1\ket{1}\\
&&\ket{\psi_R}=(e^{i\phi_R}a_0\ket{0}+a_1\ket{1})e^{i\beta+i\phi_\D+i\phi_B}
\end{eqnarray}
The phase $\phi_B$ is contributed by the bosonic exponents
carrying the charge and the spin of the quasiholes. The origin of
the phases $\phi_\Delta$, and $\phi_R$ lies in the conformal
nature of the parafermionic fields, and we explain how to
determine them below.

In case the parafermionic operator of the island is Abelian, the
coefficient $a_0$ is zero, and
$\ket{\psi_R}=\ket{1}e^{i\beta+i\phi_\D^A+i\phi_B}$ (the
superscript $A$ is added to $\phi_\D$ to indicate that the
internal state of the dot corresponds to an Abelian field). In the
case where $a_0=0$, the interference term contributed to the
backscattered current by the two partial waves, proportional to
the real part of the overlap $\bra{\psi_L}\psi_R\rangle$, is given
by
\begin{equation}\label{eq:abelian overlap}
\cos{({\rm
\arg}\bra{\psi_L}\psi_R\rangle_A})=\cos{(\beta+\phi_\D^A+\phi_B)}
\end{equation}

When the parafermionic operator of the island is non-Abelian, the
overlap is given by
\begin{equation}\label{eq:non abelian overlap}
\bra{\psi_L}\psi_R\rangle_{NA}=(e^{i\phi_R}|a_0|^2+|a_1|^2)e^{i\beta+i\phi_\D^{NA}+i\phi_B}.
\end{equation}
The ratio between the overlap in equation (\ref{eq:abelian
overlap}) and the overlap in the case where the parafermionic
operator of the island is non-Abelian is given by
\begin{equation}\label{eq:ratio}
\frac{\bra{\psi_L}\psi_R\rangle_{NA}}{\bra{\psi_L}\psi_R\rangle_A}=(e^{i\phi_R}|a_0|^2+|a_1|^2)e^{i(\phi_\D^{NA}-\phi_\D^A)}.
\end{equation}
and will determine both the relative amplitude and the phase shift
between the two cases.
\begin{itemize}
\item\textit{Determining the phases $\phi_\Delta^A,\phi_\Delta^{NA},
              \phi_R$, and $\phi_B$}.

As was mentioned before, the fusion of $n$ quasiholes results in a
parafermion multiplied by a bosonic vertex operator of the form
\begin{equation}
e^{i\left(n_{\ua}\vec{\alpha}_{\ua}+n_\da\vec{\alpha}_{\da}+n_3\vec{\alpha}_{3}\right)\cdot\vec{\varphi}}=e^{i\frac{1}{2}\left(\frac{n_{\ua}+n_{\da}+2n_{3}}{\sqrt{7}},n_{\ua}-n_{\da}\right)\cdot\vec{\varphi}}.
\end{equation}
An incoming quasihole carries, in general, a bosonic factor of the
form $e^{i\vec{\alpha}_{ex}\cdot\vec{\varphi}}$, with
$\vec{\alpha}_{ex}=(\alpha_c,\alpha_s)$ being $\alpha_\ua$,
$\alpha_\da$ or $\alpha_3$. The OPE of the two vertex operators is
\begin{eqnarray}
\lefteqn{e^{i\vec{\alpha}_{ex}\cdot\vec{\varphi}}(z)
         e^{i\frac{1}{2}\left(\frac{n_{\ua}+n_{\da}+2n_{3}}{\sqrt{7}},
            n_{\ua}-n_{\da}\right)\cdot\vec{\varphi}}(0) \
\sim} \\
&& z^{f(n_{\ua},n_{\da},n_{3})} \,
   e^{i\frac{1}{2}\left(\frac{n_{\ua}+n_{\da}+2n_{3}}{\sqrt{7}}
   +2\alpha_c,n_{\ua}-n_{\da}+2\alpha_s\right)\cdot\vec{\varphi}}(0), \nonumber
\end{eqnarray}
where
\begin{eqnarray}
f(n_{\ua},n_{\da},n_{3})=\frac{1}{2}\left[\frac{(n_\ua+n_\da+2n_3)\alpha_c}{\sqrt{7}}+(n_\ua-n_\da)\alpha_s\right]\nonumber.
\end{eqnarray}
Therefore,
\begin{equation}
\phi_B=\pi\left[\frac{(n_\ua+n_\da+2n_3)\alpha_c}{\sqrt{7}}+(n_\ua-n_\da)\alpha_s\right].
\end{equation}

 The phases $\phi_\D^{NA},\phi_\D^A$ and $\phi_R$ all depend on the
conformal dimension of the parafermionic fields. We now demonstrate
how to calculate them for a particular example (we leave out the
bosonic part of the quasihole operators). Suppose that the fusion of
$n$ quasiholes on the island can result in one of the following two
parafermions: $\psi_2$ or $\sg_\ua$. Suppose that the incoming edge
quasihole has spin up. If the quasiholes on the dot fused to the
Abelian field $\psi_2$, the OPE of the island's operator with the
incoming quasiholes is
\begin{equation}
\psi_2\cdot\sg_{\ua}\sim z^{h_{\rho}-h_{\sg}-h_{\psi}}\rho.
\end{equation}
Therefore, the phase accumulated when the quasihole encircles the
island is $\phi_\D^A=2\pi(h_{\rho}-h_{\sg}-h_{\psi})=0$.

If the $n$ quasiholes in the dot fused to yield the non-Abelian
field $\sg_\ua$, the OPE of the island's operator with the incoming
quasiholes is
\begin{equation}
\sg_\ua\sg_{\ua}\sim
z^{-2h_\sg}\mathbf{1}+z^{h_{\rho}-2h_{\sg}}\rho=
(z^{-h_\rho}\mathbf{1}+\rho)z^{h_{\rho}-2h_{\sg}}.
\end{equation}
From the above equation we conclude that $\phi_R=-2\pi
h_{\rho}=-6\pi/5$, and $\phi_\D^{NA}=2\pi(h_{\rho}-2h_\sg)=4\pi/5$.

\item\textit{Determining the values of $|a_0|^2$ and $|a_1|^2$}

The coefficients $|a_0|^2, |a_1|^2$ are the probabilities that the
quasihole encircling the island fuses with the localized bulk
quasiholes into Abelian and non-Abelian fields, respectively.

In the general context of models for non-Abelian anyons
\cite{Bonderson3,Preskill}, the probability that two uncorrelated
anyons, $a$ and $b$, fuse into a third anyon $c$ is given by (see
Ref.~(\onlinecite{Preskill}))
\begin{equation}\label{eq:prob using QD}
P(ab\rightarrow c)=N_{ab}^c\frac{d_c}{d_ad_b}
\end{equation}
where $d_{a}, d_{b}$ and $d_{c}$ are the quantum dimensions of the
anyons. The coefficient $N_{ab}^c$ is in general a non-negative
integer determined by the fusion rules, and indicates the number of
different way the anyons $a$ and $b$ can be combined to form $c$. In
our case it is either zero or one, depending on whether the fusion
rules allow $a$ and $b$ to fuse to $c$.

The quantum dimension is a parameter that controls the rate in
which the Hilbert space of a system of $n$ anyons of a certain
type grows. For a large number of such anyons, the number of
states scales as $d^{n}$. For Abelian particles $d=1$, while
non-Abelian particles have $d>1$.

The expression for the overlap between the two partial waves
$\ket{\psi_{L,R}}$ can therefore be written as follows
\begin{equation}\label{eq:non abelian overlap}
\bra{\psi_L}\psi_R\rangle_{NA}=\frac{1}{d_{bulk}d_{qh}}(e^{i\phi_R}+d_1)e^{i\beta+i\phi_\D^{NA}+i\phi_B}.
\end{equation}
Where $d_{bulk}$ and $d_{qh}$ are the quantum dimensions of the bulk
and of the edge quasihole respectively. The quantum dimension $d_1$
is that of the non-Abelian field that is the result of the fusion
product of both.  Note that this expression is exactly the
expression for the monodromy matrix element calculated in
Ref.~(\onlinecite{Bonderson3})
\begin{equation}\label{eq:monodromy}
M_{ab}=\frac{1}{d_ad_b}\sum_{c}N_{ab}^cd_ce^{2\pi i(s_c-s_a-s_b)}
\end{equation}
where $s_i$ is the scaling dimension of the relevant anyon. In our
case, $a$ represents the edge qusihole and $b$ represents the
composite made of a group of qusiholes in the bulk.

Taking for example the neutral part of the spin-up quasihole
$\sg_\ua$, we use the fusion rules to count how many states there
are for $n$ such quasiholes:
\begin{eqnarray*}
&&\sg_\ua\sg_\ua=\mathbf{1}+\rho \\
&&\sg_\ua\sg_\ua\sg_\ua=\psi_2+2\sg_\ua\\
&&\sg_\ua\sg_\ua\sg_\ua=2\mathbf{1}+3\rho\\
&&\sg_\ua\sg_\ua\sg_\ua\sg_\ua=3\psi_2+5\sg_\ua...
\end{eqnarray*}
The number of states for $n$ quasiholes is a Fibonacci number. For a
large value of $n$, this number scales as $\tau^n$, where $\tau$ is
the Golden Ratio.

Following the same considerations one finds that for the $SU(3)_2$
parafermionic fields, there are two values for the quantum
dimension: The quantum dimension of the Abelian fields
$\mathbf{1},\psi_i$ is $d=1$, while that of the non-Abelian fields
$\rho,\sigma_i$ is $d=\tau$. Therefore
\begin{equation}\label{eq:fusion probabilities}
|a_1|^2=\frac{1}{\tau},
    \quad|a_0|^2=\frac{1}{\tau^2}.
\end{equation}

\end{itemize}

Plugging the values of $\phi_R$, $\phi_\D^{NA}$ and $\phi_\D^{A}$
from the above example into Eq.~(\ref{eq:ratio}), along with the
coefficients Eq.~(\ref{eq:fusion probabilities}), and using the fact
that $\tau=2\cos(\pi/5)$, we find
\begin{eqnarray}\label{eq:damping}
\frac{\bra{\psi_L}\psi_R\rangle_{NA}}{\bra{\psi_L}\psi_R\rangle_A}=-\tau^{-2}.
\end{eqnarray}
From Eq.~(\ref{eq:damping}) we conclude that the interference term
is suppressed by a factor of $\tau^{-2}$ when the parafermionic
operator of the island is non-Abelian. 

It is straightforward to calculate the ratio Eq.~(\ref{eq:ratio})
for all other possible operators of the island. The result always
turns out the same and given by Eq.~(\ref{eq:damping}). The
interference term is therefore
\begin{eqnarray}\nonumber
&&(-\tau^{-2})^{N_{\phi}}
 \\
&& \times
\cos\left(\beta+\pi\frac{(n_\ua+n_\da+2n_3)\alpha_c}{\sqrt{7}}
    +\pi(n_\ua-n_\da)\alpha_s+\phi_\D^A\right)\nonumber\\
    &&=(-\tau^{-2})^{N_{\phi}}\cos\left(\beta+8\pi\frac{(n_\ua+n_\da+2n_3)\alpha_c}{\sqrt{7}}\right).\label{eq:inter term su3}
\end{eqnarray}
Here
$\phi_\D^A=\pi(\lambda_1^{bulk}\lambda_2^{qh}+\lambda_2^{{bulk}}\lambda_1^{qh})$
is $0$ or $\pi$ (up to $2\pi$) depending on the specific Abelian
field that may result from fusing $n$ bulk quasiholes. The vector
$(\lambda_1,\lambda_2)$ is given for each operator in equation
(\ref{eq:SU(3)2 pararefmions}), $\lambda_i^{bulk}$ are the
components of the vector that corresponds to the fusion product of
the bulk quasiholes, and $\lambda_i^{qh}$ are the components of
the vector that corresponds to the quasihole that tunnels across
the point-contact. The second line of the equation is obtained
from the first with some algebra. The number $N_{\phi}$ is either
$0$ or $1$ depending on whether the fusion product of the
quasiholes on the dot was an Abelian or non-Abelian field.
Eq.~(\ref{eq:inter term su3}) shows that the damping factor that
multiplies the interference term is the same as the one predicted
to appear for the $k=3$ RR state. This is a manifestation of the
so-called "level-rank duality" between $SU(3)_2$ and $SU(2)_3$,
see Ref.~(\onlinecite{StateCounting}). This result agrees with the
one calculated using the modular S-matrix\cite{ArdonneSmatrix} in
a similar fashion to the calculation presented in
Ref.~(\onlinecite{Bonderson2}) for lowest order interference in
the RR series.

The above result is calculated without making any assumptions on the
nature of the edge qusiholes that tunnels across the point contact,
and is therefore general. However, it is expected that the most
prominent contribution to these tunneling events will come from the
quasihole with the lowest conformal dimension. For that reason, one
may expect mostly spinless quasiholes to tunnel across the junction.
The interference term for these quasiparticle is given by equation
(\ref{eq:inter term su3}) with $\alpha_c=1/\sqrt{7}$ and
$\alpha_s=0$.

\section{Summary}\label{sec:summary}
In this work we studied transport through a device known as the
two point-contact (or the Fabry-Perot) interferometer formed using
edge states of non-Abelian quantum Hall phases. We have focused on
the RR series at filling factor $\nu=2+\frac{k}{k+2}$, and on one
of the spin singlet states that may be appropriate to describe the
plateau at filling factor $4/7$.

For the limit of strong quasihole backscattering at the two point
contacts, where the topmost partially filled level forms a closed
trajectory around the "island" formed between the point contacts,
transport is characterized by a series of Coulomb blockade peaks in
the conductance as a function of the area enclosed by this
trajectory.

For the RR states we explained in details the results obtained in
Ref.~(\onlinecite{ilan}), and analyzed the effect of relaxation of
the edge states of the parafermionic theory into states with lower
energy by adjusting the fusion channel of the bulk quasiholes. We
found that this relaxation, mediated by neutral fields of the
parafermion theory flowing between the bulk and the edge, modifies
the spacing between peaks. Without relaxation, the Coulomb
blockade pattern predicted~\cite{ilan} is a bunching of peaks into
groups of $n\bmod{k}$ and $k-n\bmod{k}$ peaks (where $n$ is the
number of quasiholes localized in the bulk). Therefore the
periodicity of the peak structure is $k$, unless $k$ is even and
$n=k/2$, in which case the periodicity is $k/2$ (assuming, for
convenience, that the initial number of electrons in the dot was
divisible by $k$). When relaxation is introduced between
consecutive electron-tunneling events, this structure changes such
that the number of bulk quasiholes no longer influences the
periodicity of the entire structure. The periodicity is $k$ for
odd $k$ and $k/2$ for even $k$. For the $k=2$ Moore-Read state, we
predict that in the presence of relaxation of Majorana fermions
from the edge of the dot into the bulk, the even-odd effect
predicted in Ref.~(\onlinecite{SternHalperin}) will not be
observed. Therefore, for that state relaxation eliminates the
unique signature for non-Abelian statistics and clustering.

We would like to stress the difference between two time scales
relevant for the experiments we discuss here, that of relaxation
of the edge states which is discussed in the Coulomb blockade
context, and that of quasihole exchange between the bulk and the
edge, which is neglected altogether. The two scales are related to
different mechanisms and are generally very different. The first
involves an exchange of neutral particles between the edge and the
bulk, and the second involves tunneling of a charged particle
(with both non-zero $\mathbb{Z}_k$ and electric charge). The
charging energy involved in the latter is expected to make this
time scale much longer than the former: Hopping of quasiholes
between the bulk and the edge involves energy scales of the order
of the bulk gap, while relaxation of edge states via the exchange
of neutral modes involves an energy which is much smaller and
scales as $L^{-1}$, where $L$ is the perimeter of the dot. Little
can be said about these time scales quantitatively, since they
crucially depend on disorder in the sample, and in a way that is
not presently understood.

The microscopic mechanism which induces relaxation is beyond the
scope of this paper. Since at present there are many fundamental
aspects of the microscopic theory of non-Abelian quantum Hall
states that are not well understood, we investigate the
implication of such a mechanism on a phenomenological level and
provide a prediction that follows form it. While not giving
quantitative estimates, our paper brings to light the possible
effect of the time scales related to these mechanisms on
experiments in non-Abelian quantum Hall states.

A few particular sentences are in order with respect to the even-odd
effect at $\nu=5/2$, both because it is at present the closest to
experimental realization and because it stands out as a special case
in our analysis. We have found, that in the presence of an inelastic
bulk-edge coupling that allows the system to be at its ground state
all throughout the experiment, and in the presence of bulk
quasiholes, the edge mode stays unpopulated in the ground state for
any number of electrons in the dot. Thus, under this condition no
even-odd effect of the Coulomb blockade peaks will be observed.
However, we stress that the situation is different when the bulk
edge coupling is elastic, as considered for example in
(\onlinecite{BulkEdgeCoupling,OverboschWen}). In that case, the
spectrum of the combined bulk-edge system is shifted by the
coupling. The coupling introduces a time scale for a Majorana
fermion to tunnel back and forth between the edge and the bulk. As
long as this time scale is much longer than the time at which the
Majorana fermion encircles the dot, $L/v_n$, the bulk-edge coupling
does not significantly affect the spectrum, and an even-odd effect
is to be observed.

For the spin singlet state at $\nu=4/7$ we have mapped the location
of the Coulomb blockade peaks as a function of the area of the dot,
and showed that it follows a periodicity of $2$ or $4$. Moreover,
although the modulations from equal spacings change when relaxation
is introduced, the periodicity remains unaffected.

One should also note another crucial difference between the RR
states and the spin-singlet state, that may be observed in such an
experiment. The extent to which peaks are bunched in the RR state
is determined by the ratio of the velocity of the neutral modes to
that of the charged modes, $v_n/v_c$, as they propagate along the
edge. These velocities are hard to determine; in the absence of
Landau level mixing and for an infinitely sharp confining
potential, dimensional analysis determines that the
interaction-induced velocity is, up to a dimensionless number,
$e^2/\epsilon\hbar$, with $\epsilon$ being the dielectric
constant. The dimensionless number may be evaluated only
numerically, and one expects it to be larger for a charged mode
than for a neutral one. Indeed, recent numerical
studies\cite{PffafianNumerics} for $\nu=5/2$ predict that the two
velocities differ by an order of magnitude. If that trend persists
to RR states of higher $k$, then the effect of bunching in the RR
states may be difficult to observe in experiment. In contrast, the
size of the modulations in the spacing between Coulomb blockade
peaks for the spin-singlet state is determined by two
contributions, the first is $v_n/v_c$ which may be small for this
state as well, and the second is $v_s/v_c$, the ratio between the
velocity of the spin and charge excitations. If this ratio is not
small, then these modulations may still be observed. Note,
however, that in the extreme case of $v_s>>v_c$, the periodicity
in some cases may seem to reduce from $4$ to $2$ (see Table
\ref{tab:tunneling sequences}). Moreover, as we point out below,
these modulations in the spacings due to the spin degree of
freedom, may not be unique for a non-Abelian state.

While it was shown before that Coulomb blockade peaks are equally
spaced for the Laughlin states, the case of other Abelian quantum
Hall states was not discussed (when the parameter varied is the
area of the dot, which keeps the number of bulk quasiholes fixed).
One may expect that for Abelian hierarchy states peaks may not be
equally spaced, since the edge theory describing these states is
made of several chiral boson theories. However, we expect that the
structure and its periodicity in particular will not show
dependence on the number bulk quasiholes. Thus, several
measurements at slightly different values of the magnetic field,
should in principle distinguish an Abelian hierarchy state from a
non-Abelian one. For the spin-singlet state at $\nu=4/7$ it may be
more difficult to distinguish such Abelian states from non-Abelian
ones, since modulation may still occur for $v_n=0$ due to spin.
However, the periodicity of the peak structure predicted is
different for negligible and non-negligible values of $v_n/v_c$,
and this will ultimately distinguish an Abelian state from a
non-Abelian one. The case of other Abelian states may be slightly
more complicated. For example, the $331$ Halperin state may show
bunching as well as a dependence on the number of
quasiholes.\cite{foot2}

Finally, we considered lowest order interference effects in the
Fabry-Perot interferometer, and calculated the form of the
backscattered quasihole current through this device. For the
non-Abelian spin-singlet state we have found that interference
will be suppressed by a damping factor that is equal to the one
obtained for the RR state with $k=3$ in
Ref.~(\onlinecite{Bonderson2}). The phase of the interference term
has a rich structure, and is influenced by the number of
quasiholes localized in the bulk, their total charge and spin, and
the spin of the interfering quasihole.

\acknowledgements{We would like to thank Kirill Shtengel for useful
discussions. We thank the support of the Einstein Minerva center for
theoretical physics. RI, EG and AS acknowledge support from the
US-Israel Binational Science Foundation and the Minerva foundation.
KS is supported by the foundation FOM of the Netherlands and by the
INSTANS programme of the ESF.} \hspace{-0.5in}

\begin{table*}[b]
\begin{center}
\begin{tabular}{|c|c|l|l|c|}
  \hline
  \text{Highest} &\quad  \quad&\quad\quad&\quad\quad&\quad  \quad\\
  \text{weight} &\quad \text{Tunneling} \quad&\quad\quad&\quad\quad&\quad  \quad\\
  \text{state} &\quad \text{sequence} \quad&\quad\text{Energy spectrum}\quad&\quad\text{Spacings}\quad&\text{Periodicity}\\
  \hline
  $\ket{0}$&\ \text{for}$\quad v_n/v_s>1$  & $E_0=E_c(N_e)$ & $\Delta S_1=\frac{e}{n_0}\left(1+\nu\frac{v_s}{4v_c}-\nu\frac{v_n}{v_c}\right)$ & $4$\\
  $\ $&$\ \mathbf{1}\stackrel{\psi_1}\longrightarrow\psi_1\stackrel{\psi_1}\longrightarrow\mathbf{1}$  & $E_1=E_c(N_e+1)+\frac{\pi v_n}{ L}+\frac{\pi v_s}{4L}$ & $\Delta S_2=\frac{e}{n_0}\left(1-\frac{3}{4}\nu\frac{v_s}{v_c}+\nu\frac{v_n}{v_c}\right)$ & \\
  $\ $&$\quad\quad\stackrel{\psi_2}\longrightarrow\psi_2\stackrel{\psi_2}\longrightarrow\mathbf{1} $  & $E_2=E_c(N_e+2)+\frac{\pi v_s}{L}$ & $\Delta S_3=\frac{e}{n_0}\left(1+\nu\frac{v_s}{4v_c}-\nu\frac{v_n}{v_c}\right)$ & \\
  $\ $&$\ $  & $E_3=E_c(N_e+3)+\frac{\pi v_n}{ L}+\frac{\pi v_s}{4L}$ & $\Delta S_1=\frac{e}{n_0}\left(1+\nu\frac{v_s}{4v_c}+\nu\frac{v_n}{v_c}\right)$ & \\\cline{2-5}
  $\ $&\ \text{for}$\quad v_n/v_s<1$  & $ E_0=E_c(N_e)$ & $\Delta S_1=\frac{e}{n_0}\left(1-\nu\frac{v_s}{4v_c}-\frac{1}{2}\nu\frac{v_n}{v_c}\right)$ & $4$ \\
  $\ $&$\ \mathbf{1}\stackrel{\psi_1}\longrightarrow\psi_1\stackrel{\psi_2}\longrightarrow\psi_{12}$  & $E_1=E_c(N_e+1)+\frac{\pi v_n}{ L}+\frac{\pi v_s}{4L}$ & $\Delta S_2=\frac{e}{n_0}\left(1+\nu\frac{v_s}{4v_c}\right)$ & $\ $\\
  $\ $&$\quad\quad\stackrel{\psi_{1(2)}}\longrightarrow\psi_{2(1)}\stackrel{\psi_{2(1)}}\longrightarrow\mathbf{1} $  & $E_2=E_c(N_e+2)+\frac{\pi v_n}{L}$ & $\Delta S_3=\frac{e}{n_0}\left(1-\nu\frac{v_s}{4v_c}-\frac{1}{2}\nu\frac{v_n}{v_c}\right)$ & $\ $\\
  $\ $&$\ $  & $E_3=E_c(N_e+3)+\frac{\pi v_n}{ L}+\frac{\pi v_s}{4L}$ & $\Delta S_4=\frac{e}{n_0}\left(1+\nu\frac{v_s}{4v_c}+\nu\frac{v_n}{v_c}\right)$ & $\ $\\\cline{1-5}
  $\ket{\sg_\ua}$&$\ \sigma_\ua\stackrel{\psi_1}\longrightarrow\sigma_3\stackrel{\psi_2}\longrightarrow\sigma_\da $& $E_0=E_c(N_e)+\frac{1}{10}\frac{2\pi v_n}{ L}$ & $\Delta S_1=\Delta S_3=\frac{e}{n_0}\left(1-\nu\frac{v_s}{4v_c}\right)$ & $2 $ \\
  $\quad$ & $\quad\quad\stackrel{\psi_2}\longrightarrow\sigma_3\stackrel{\psi_1}\longrightarrow\sigma_\ua$ &  $E_1=E_c(N_e+1)+\frac{1}{10}\frac{2\pi v_n}{ L}+\frac{\pi v_s}{4L}$ & $\Delta S_2=\Delta S_4=\frac{e}{n_0}\left(1+\nu\frac{v_s}{4v_c}\right) $  &   \\
  $\quad$ & $\quad$ &  $E_2=E_c(N_e+2)+\frac{1}{10}\frac{2\pi v_n}{L}$ & $\ $  &  $\ $ \\
  $\quad$ & $\quad$ &  $E_3=E_c(N_e+3)+\frac{1}{10}\frac{2\pi v_n}{L}+\frac{\pi v_s}{4L}$ &   & \\\cline{1-5}
  $\ket{\sg_\da}$&$\ \sg_\da\stackrel{\psi_2}\longrightarrow\sigma_3\stackrel{\psi_1}\longrightarrow\sigma_\ua$ & \text{The same as above} & \text{The same as above} & $2$ \\
  $\ $&$\quad\quad\stackrel{\psi_1}\longrightarrow\sigma_3\stackrel{\psi_2}\longrightarrow\sigma_\da$ &  & &  \\\cline{1-5}

  $\ket{\sg_3}$&\ \text{for}$\quad v_n/v_s>1$  & $E_0=E_c(N_e)+\frac{1}{10}\frac{2\pi v_n}{ L}$ & $\Delta S_1=\frac{e}{n_0}\left(1+\nu\frac{v_s}{4v_c}\right)$& $4$\\
  $\ $&$\ \sigma_3\stackrel{\psi_1}\longrightarrow\sigma_\ua\stackrel{\psi_1}\longrightarrow\sigma_3$  & $E_1=E_c(N_e+1)+\frac{1}{10}\frac{2\pi v_n}{L}+\frac{\pi v_s}{4L}$ & $\Delta S_2=\frac{e}{n_0}\left(1-\frac{3}{4}\nu\frac{v_s}{v_c}\right)$ & \\
  $\ $&$\quad\quad\stackrel{\psi_2}\longrightarrow\sigma_\da\stackrel{\psi_2}\longrightarrow\sigma_3 $  & $E_2=E_c(N_e+2)+\frac{1}{10}\frac{2\pi v_n}{ L}+\frac{\pi v_s}{L}$ & $\Delta S_3=\frac{e}{n_0}\left(1+\nu\frac{v_s}{4v_c}\right)$ & \\
  $\ $&$\ $  & $E_3=E_c(N_e+3)+\frac{1}{10}\frac{2\pi v_n}{ L}+\frac{\pi v_s}{4L}$ & $\Delta S_4=\frac{e}{n_0}\left(1+\nu\frac{v_s}{4v_c}\right)$ & \\\cline{2-5}
  $\ $&\ \text{for}$\quad v_n/v_s<1$  & $ E_0=E_c(N_e)+\frac{1}{10}\frac{2\pi v_n}{ L}$ & $\Delta S_1=\frac{e}{n_0}\left(1-\nu\frac{v_s}{4v_c}+\frac{1}{2}\nu\frac{v_n}{v_c}\right)$ & $4$ \\
  $\ $&$\ \sigma_3\stackrel{\psi_1}\longrightarrow\sigma_\ua\stackrel{\psi_2}\longrightarrow\rho$  & $E_1=E_c(N_e+1)+\frac{1}{10}\frac{2\pi v_n}{L}+\frac{\pi v_s}{4L}$ & $\Delta S_2=\frac{e}{n_0}\left(1+\nu\frac{v_s}{4v_c}-\nu\frac{v_n}{v_c}\right)$ & $\ $\\
  $\ $&$\quad\quad\stackrel{\psi_1}\longrightarrow\sigma_\da\stackrel{\psi_2}\longrightarrow\sigma_3 $  & $E_2=E_c(N_e+2)+\frac{3}{5}\frac{2\pi v_n}{ L}$ & $\Delta S_3=\frac{e}{n_0}\left(1-\nu\frac{v_s}{4v_c}+\frac{1}{2}\nu\frac{v_n}{v_c}\right)$ & $\ $\\
  $\ $&$\ $  & $E_3=E_c(N_e+3)+\frac{1}{10}\frac{2\pi v_n}{ L}+\frac{\pi v_s}{4L}$ & $\Delta S_4=\frac{e}{n_0}\left(1+\nu\frac{v_s}{4v_c}\right)$ & $\ $\\\cline{1-5}
  \end{tabular}

\caption{Tunneling sequence for charge transport through a quantum
dot in the $\nu=4/7$ non-Abelian spin-singlet quantum Hall state.
In case the highest weight state is $\ket{0}$, replacing $\psi_1$
with $\psi_2$ and vice versa everywhere in the sequence yields the
same set of energies.} \label{tab:tunneling sequences}
\end{center}
\end{table*}


\begin{thebibliography}{36}
\bibitem[{\citenamefont{foot1}()}]{foot1}
One should single out the cases where $n<k$ and $s\neq 0$. In such
case, a fusion channel leading to the $\mathbb{Z}_k$ charge equal
to $[2s-n]_k$ giving minimal energy may not exist, leading to
modifications in the predictions.
\end{thebibliography}

\end{document}